\documentclass[aps,prl,reprint,twocolumn,superscriptaddress,floatfix,nofootinbib,longbibliography]{revtex4-1}
\usepackage{graphicx,amsmath,amsfonts,amssymb,amsthm,xr}

\usepackage{epsfig,amsmath,amssymb,color,dsfont,upgreek,physics}
\usepackage{mathrsfs}
\usepackage{mathtools}
\usepackage{bbold}
\usepackage{comment}
\usepackage{float}
\usepackage{braket}

\usepackage[bookmarks=true,colorlinks,linkcolor=OrangeRed,urlcolor=NavyBlue,citecolor=RoyalBlue]{hyperref}
\usepackage[dvipsnames]{xcolor}

\usepackage[normalem]{ulem}
\def\d{\mathrm d}

\begin{document}

\title{Embedding semiclassical periodic orbits into chaotic many-body Hamiltonians}

\author{Andrew Hallam}

\affiliation{School of Physics and Astronomy, University of Leeds, Leeds LS2 9JT, UK}

\author{Jean-Yves Desaules}
\affiliation{School of Physics and Astronomy, University of Leeds, Leeds LS2 9JT, UK}

\author{Zlatko Papi\'c}
\affiliation{School of Physics and Astronomy, University of Leeds, Leeds LS2 9JT, UK}

\begin{abstract}
Protecting coherent quantum dynamics from chaotic environment is key to realizations of fragile many-body phenomena and their applications in quantum technology. We present a general construction that embeds a desired periodic orbit into a family of non-integrable many-body Hamiltonians, whose dynamics is otherwise chaotic. Our construction is based on time dependent variational principle that projects quantum dynamics onto a manifold of low-entangled states, and it complements earlier approaches for embedding non-thermal eigenstates, known as quantum many-body scars, into thermalizing spectra. By designing terms that suppress ``leakage" of the dynamics outside the variational manifold, we engineer families of Floquet models that host exact scarred dynamics, as we illustrate using a driven Affleck-Kennedy-Lieb-Tasaki model and a recent experimental realization of scars in a dimerized superconducting qubit chain.   
\end{abstract}

\maketitle

{\em Introduction.---}The dynamics of non-integrable quantum many-body systems typically gives rise to rapid thermalization and scrambling of information. These hallmarks of quantum ergodicity can be traced to the properties of the system's mid-spectrum eigenstates, which are generally highly entangled and obey the Eigenstate Thermalization Hypothesis (ETH)~\cite{DeutschETH, SrednickiETH}. In recent years, there has been a flurry of activity aimed at understanding the conditions for \emph{weak} breaking of the ETH to emerge, in particular by devising ways of embedding non-thermalizing eigenstates into otherwise chaotic many-body spectra~\cite{Serbyn2021, MoudgalyaReview, ChandranReview}. These eigenstates, referred to as quantum many-body scars (QMBSs), have been identified in prominent models of quantum magnets, such as the Affleck-Kennedy-Lieb-Tasaki (AKLT) model~\cite{AKLT, Arovas1989, BernevigEnt}, and in Rydberg atom quantum simulators~\cite{Turner2017, Turner2018b}, where their signatures were first observed in quench experiments~\cite{Bernien2017}. Potential applications of QMBSs have been explored in the context of controlling  quantum-information dynamics in complex systems~\cite{Bluvstein2021} and for quantum metrology~\cite{Dooley2021, DesaulesQFI, Dooley2022}.

Despite much interest in weak ergodicity breaking phenomena in different experimental platforms~\cite{Jepsen2021, GuoXian2022, Zhang2022}, the origin of QMBSs remains the subject of ongoing investigation. In much of theoretical work, QMBSs are studied by algebraic constructions of ergodicity-breaking eigenstates. In particular, the local projector approach by Shiraishi and Mori~\cite{ShiraishiMori} embeds a few non-thermal eigenstates into the spectrum of a non-integrable Hamiltonian. Other approaches construct families of eigenstates, representing condensates of quasiparticles evenly spaced in energy~\cite{BernevigEnt, Iadecola2019} . 
More recent proposals aim to unify these different constructions into a single framework~\cite{MotrunichTowers, Dea2020, Pakrouski2020, Moudgalya2022}. All these approaches, however, differ dramatically from the case of single-particle scars in quantum billiards, which are understood as quantum remnants of classical, unstable periodic orbits~\cite{Heller84, HellerLesHouches, Bogomolny1988, Berry1989Scars}. Nevertheless, for QMBSs observed in Rydberg atom experiments~\cite{Bernien2017}, the eigenstate constructions~\cite{Turner2018b, Iadecola2019, lin2018exact, Choi2018, Bull2020} were shown to be in harmony with a  semiclassical limit of the dynamics, developed by Ho~\emph{et al.}~\cite{wenwei18TDVPscar},  which identified a periodic orbit in the many-body Hilbert space that underpins the coherent QMBS dynamics. The notion of a semiclassical limit, introduced in Ref.~\cite{wenwei18TDVPscar} and adopted in this paper, is based on projecting quantum dynamics to a variational manifold spanned by states with low entanglement.

\begin{figure}
\centering
    \includegraphics[width=1.0\linewidth]{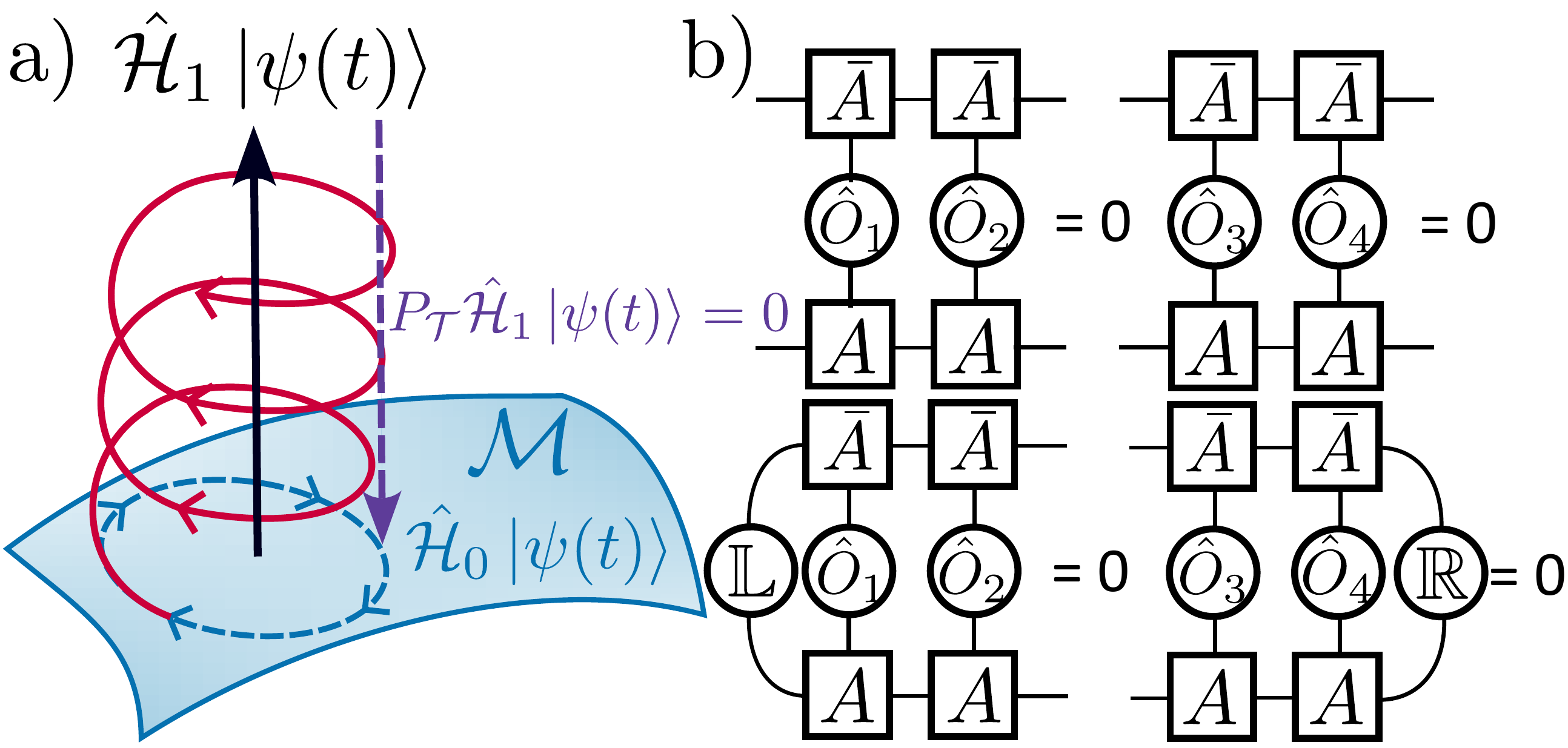}
     \caption{(a) A periodic orbit can be embedded into a chaotic many-body  Hamiltonian by decomposing the latter into two terms: $\hat{\mathcal{H}}_0$, which generates a periodic orbit in a semiclassical manifold $\mathcal{M}$, and $\hat{\mathcal{H}}_1$ that vanishes as the semiclassical limit is taken.   (b) For an MPS manifold, $\hat{\mathcal{H}}_1$ takes a simplified form. The generic conditions for a four-site local $\hat{\mathcal{H}}_1$, valid for any system size and choice of boundary conditions, are illustrated (top). These conditions can be weakened for translation invariant MPS in the thermodynamic limit (bottom).
     }
         \label{fig:Fig1}
\end{figure}

In this work, we introduce a systematic method for embedding a desired periodic trajectory into the dynamics generated by a chaotic many-body Hamiltonian. 
The latter is understood to obey the ETH, apart from a vanishing fraction of states in the thermodynamic limit. Our method is based on decomposing the Hamiltonian into a component that generates an exact periodic orbit and a second component that vanishes upon taking the semiclassical limit. 
Thus, within a suitably-defined semiclassical manifold, the projected dynamics is a periodic oscillation.  However, the dynamics of the full model may deviate from the projection to the manifold and this deviation is quantified by the so-called quantum leakage~\cite{wenwei18TDVPscar, Michailidis2020}. Using the quantum leakage, we introduce driving terms to the model that cancel the distinction between the semiclassical and quantum dynamics, schematically depicted in Fig.~\ref{fig:Fig1}, which results in exact Floquet QMBSs. We demonstrate the utility of our approach using the AKLT model~\cite{AKLT, BernevigEnt} and a recent superconducting circuit realization of QMBS~\cite{Zhang2022} based on the Su-Schrieffer-Heeger (SSH) model~\cite{SSH}.  

{\em Time-dependent variational principle (TDVP).---}To avoid the exponential complexity of many-body quantum systems, the TDVP method~\cite{Dirac1930, kramer1981geometry} approximately solves the time-dependent Schr\"odinger equation (TDSE) by projecting it onto a manifold $\cal M$ spanned  by ansatz wave functions that capture the most important features of the dynamics.  In this work we focus on one-dimensional lattice systems with a $d$-dimensional Hilbert space on each site, where $\cal M$ is spanned by wave functions $\ket{\psi(\{z_n\})}$, parameterized by a complex variable $z_n$ for each site $n$. For example, in a simple manifold describing product states of spins-1/2, one can think of $z_n$ parameterizing the orientation of each spin on the Bloch sphere. However, as pointed out by Haegeman~\emph{et al.}~\cite{Haegeman}, a much larger class of dynamical behaviors can be described if we allow $\cal M$ to contain entangled states such as  matrix product states (MPS)~\cite{PerezGarcia}. For MPS, the variables $z_n$ on each site are $\chi {\times} \chi$ dimensional matrices, $A^s_n$, labelled by a local basis vector $s=1,2,\ldots d$ and  site index $n$. Increasing $\chi$ increases the power of the ansatz, representing states with larger amounts of entanglement between sites.  

The time evolution within $\cal M$ is given by~\cite{kramer1981geometry}
\begin{equation}
\label{eq:TDVP}
i\frac{\d}{\d t}\ket{\psi(\{z_n\})}=P_{\mathcal{T}}\hat{\mathcal{H}}\ket{\psi(\{z_n\})},
\end{equation}
where 
$P_{\mathcal{T}} = \sum_n \ket{\partial_{z_n}\psi(\{z_n\})}g^{-1}_{z_n\bar{z}^\prime_n}\bra{\partial_{\bar{z}^\prime_n}\psi(\{\bar{z}^\prime_n\})}$ is a projector onto the tangent-space of $\cal M$ at the point $\ket{\psi(\{z_n\})}$. $g^{-1}_{\bar{z}_n z^\prime_n}$ is the inverse of the metric tensor of $\cal M$, $g_{\bar{z}_n z^\prime_n}=\braket{\partial_{\bar{z}_n}\psi(\{\bar{z}_n\})|\partial_{z^\prime_n}\psi(\{z^\prime_n\})}$.
Due to the tangent-space projectors dependence on $\{z_n\}$, the TDVP dynamics typically deviate from that generated by the TDSE, becoming nonlinear. When $\cal M$ is a so-called K{\"a}hler manifold, it is a classical dynamical phase space with the TDVP equations being the corresponding Hamilton equations \cite{huybrechts2005complex,Haegeman2014geometry}. Additionally, when the states in $\cal M$ form an overcomplete basis, a Feynman path integral over $ \cal M$ can be constructed \cite{green2016feynman}. The TDVP equations then correspond to the Euler-Lagrange equations of the path integral. 

{\em Semiclassical limit.---}The deviation between TDVP and TDSE can be characterized using  \emph{quantum leakage} $\Gamma$~\cite{wenwei18TDVPscar}. The leakage is the norm of the difference between the full and approximate time-evolved wave functions, integrated around the orbit:
\begin{equation}
\Gamma=\frac{1}{T}\oint||(\mathbb{1}-P_{\mathcal{T}})\hat{\mathcal{H}}\ket{\psi(\{z_n(t)\})}|| \; \d t.
\end{equation}
Typically, $\Gamma^2$ is extensive, i.e., asymptotically proportional to the system size $N$. By constraining the complexity of $\cal M$, the TDVP approach allows one to effectively \emph{define} a semiclassical limit of the full quantum dynamics~\cite{wenwei18TDVPscar}: provided the full quantum dynamics are well approximated within the manifold, i.e., $\Gamma \ll \sqrt{N}$, and $\cal M$ is spanned by low bond dimension MPS states, we will refer to such dynamics as ``semiclassical". Note that this definition admits a semiclassical limit that includes (short-range) quantum correlations, which is essential, e.g., for capturing the behavior of constrained systems~\cite{wenwei18TDVPscar}.

{\em Orbit embedding conditions.---}We now focus on  Hamiltonians $\hat{\mathcal{H}}_0$ that possess a periodic orbit from a certain initial state, $\ket{\psi(t)}=\ket{\psi(t+T)}$, for which it is possible to find a low-dimensional $\mathcal{M}$ that exactly captures the dynamics. Suppose the Hamiltonian is then perturbed, $\hat{\mathcal{H}}_0\rightarrow \hat{\mathcal{H}}_0+\hat{\mathcal{H}}_1$, so that the TDSE is altered, but  Eq.~(\ref{eq:TDVP}) is not. A Hamiltonian that satisfies the following conditions along the trajectory will retain a semiclassical periodic orbit, despite its quantum dynamics being altered:
\begin{eqnarray}
\label{eq:cond0}
P_{\mathcal{T}}\hat{\mathcal{H}}_0\ket{\psi(\{z_n\})} = \hat{\mathcal{H}}_0\ket{\psi(\{z_n\})}, \\
\label{eq:cond1}
\left[\hat{\mathcal{H}}_1-\bra{\psi(\{\bar{z}_n\})}\hat{\mathcal{H}}_1\ket{\psi(\{z_n\})}\right]\ket{\psi(\{z_n\})} \neq 0, \\
\label{eq:cond2}
P_{\mathcal{T}}\hat{\mathcal{H}}_1\ket{\psi(\{z_n\})} = 0.
\end{eqnarray}
These conditions are illustrated in Fig.~\ref{fig:Fig1}(a).  Eq.~(\ref{eq:cond1}) and Eq.~(\ref{eq:cond2}) require that $\ket{\psi(\{z_n\})}$ is a fixed point of the TDVP equations with respect to $\hat{\mathcal{H}}_1$, while not being an eigenstate. For Hamiltonians that satisfy the conditions Eq.~(\ref{eq:cond0})-(\ref{eq:cond2}), the leakage can be simplified:
\begin{equation}
\label{eq:leakageform}
\Gamma=\frac{1}{T}\oint||\hat{\mathcal{H}}_1\ket{\psi(\{z_n(t)\})}|| \; \d t.
\end{equation}

{\em Periodic orbit for MPS.---}
Restricting our discussion to MPS as a variational ansatz, we can be more precise about the form $\hat{\mathcal{H}}_1$ must take in order to satisfy Eq.~(\ref{eq:cond2}). 

Suppose $\hat{\mathcal{H}}_1$ can be written as a sum of $2K$-local operators, $\hat{\mathcal{H}}_1=\sum_n \hat{O}^{n}_1\otimes \hat{O}^{n+1}_2 \otimes \cdots \otimes \hat{O}^{n+2K-1}_{2K}$, where $\hat{O}^{n}_k$ is the $k$-th type of local term in $\hat{\mathcal{H}}_1$ acting on site $n$. For evaluating correlation functions with MPS, it is useful to introduce the MPS transfer matrix, $\mathbb{E}_n(\hat{O}_k)=\sum_{s,\bar{s}^\prime}\bar{A}^{s^\prime}_n \hat{O}^n_{k,s,\bar{s}^\prime} A^s_n$~\cite{CiracRMP}.
In the Supplementary Material~\cite{SOM}, we prove that Eq.~(\ref{eq:cond2}) is satisfied for any system size and choice of boundary conditions if we impose:
\begin{equation}
\label{eq:MPScond}
 \prod^{K}_{k=1} \mathbb{E}_{n+k-1}(\hat{O}_k)=0, \quad \prod^{2K}_{k=K+1}  \mathbb{E}_{n+k-1}(\hat{O}_k)=0.
\end{equation}
In the thermodynamic limit, these conditions can be significantly weakened. Let us assume that $A^s_n$ is site-independent and $\mathbb{E}_n(\mathbb{1})$ possesses a unique dominant left and right eigenvector, $(\mathbb{L}|$ and $|\mathbb{R})$, respectively. In this case, Eq.~(\ref{eq:TDVP}) will always begin with $(\mathbb{L}|$ and end with $|\mathbb{R})$, so  Eq.~(\ref{eq:cond2}) is satisfied, provided 
\begin{equation}
\label{eq:MPScond_inf}
(\mathbb{L}|\prod^{K}_{k=1} \mathbb{E}_{n+k-1}(\hat{O}_k)=0, \quad \prod^{2K}_{k=K+1} \mathbb{E}_{n+k-1}(\hat{O}_k)|\mathbb{R})=0.
\end{equation}
These conditions are illustrated in Fig.~\ref{fig:Fig1}(b) and below we demonstrate how they can be used to construct families of models that share the same periodic orbit, using SSH and AKLT chains as examples. We note that the above assumption about the form of $\hat{\mathcal{H}}_1$ can be straightforwardly lifted for Hamiltonians that are sums of local operators or feature long-range interactions.

\begin{figure}[t]
    \centering %
  \includegraphics[width=1.0\linewidth]{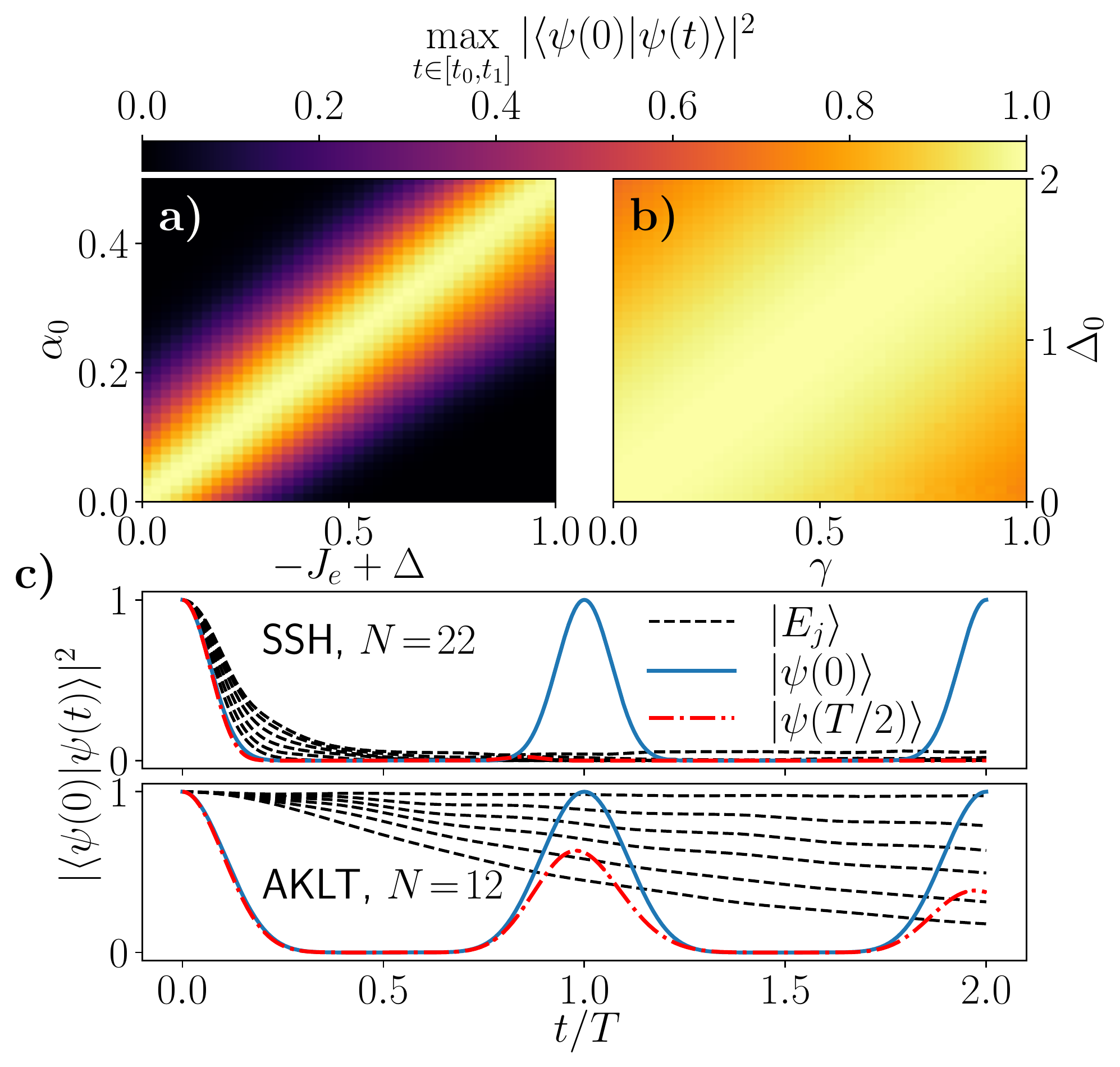}
\caption{(a) Maximum fidelity revival (between times $t_0=1$ and $t_1=2\pi$) for the driven SSH model in Eq.~(\ref{eq:SSH2}) with $\alpha(t)=-\alpha_0\sin(2J_o t)$. The system size $N=80$ and coupling $J_e=2/3$ are fixed, while $\alpha_0$ and $\Delta$ are varied.  (b) Maximum fidelity revival (between $t_0=0.5$ and $t_1=\pi$) for the driven AKLT model, Eq.~(\ref{eq:AKLTpert}), with $\Delta(t)=\Delta_0\sin(\epsilon t)$. Data is for the system size $N=50$, varying $\Delta_0$ and $\gamma$. In both (a)-(b), data is obtained using numerical implementation of TDVP with bond dimension $\chi=64$.
 (c): The scarred eigenstates $\ket{E_j}$ of the static model are destroyed by the Floquet operator, only the periodic orbit $\ket{\psi(0)}$ is preserved. Also shown is the periodic orbit shifted by $T/2$. Data for both models in (c) is obtained via  exact diagonalization.  
 }
\label{fig:Fig2}
\end{figure}

{\em SSH chain.---}We now apply our approach to the dimerized SSH model of polyacetylene~\cite{SSH,DeLeseleuc2019},
\begin{equation}
\hat{\mathcal{H}}_\mathrm{SSH} = \hspace{-0.25cm}\sum^{N/2-1}_{n=0}\hspace{-0.15cm} J_o \sigma^+_{2n+1}\sigma^-_{2n+2}+\hspace{-0.25cm}\sum^{N/2-2}_{n=0}\hspace{-0.15cm}J_e \sigma^+_{2n+2}\sigma^-_{2n+3} + \mathrm{h.c.}, \label{eq:SSH}
\end{equation}
where $\sigma^\pm$ denote the Pauli raising and lowering spin operators, $J_o$ and $J_e$ are the hopping amplitudes on the odd and even sublattice, respectively, and we have assumed open boundary conditions. In Ref.~\cite{Zhang2022}, the SSH chain was used as a starting point to realize QMBS dynamics on a superconducting quantum processor when additional couplings between sites are added to break the integrability.
In the absence of inter-dimer couplings, $J_e{=}0$, the state $\ket{\psi(0)}= \ket{10011001\cdots}$, i.e., with dimers alternating between $10$ and $01$ local states,  undergoes free precession,  with frequency $2J_o$. The oscillations are no longer perfect at  $J_e\approx 2J_o/3$ and instead exhibit a decaying envelope~\cite{Zhang2022}. It was found that a translation invariant next-next nearest neighbor hopping enhances the QMBS oscillations. Indeed, such a term reduces leakage from the scarred subspace, but it does not lead to its total suppression~\cite{SOM}. However, using the above approach, we can identify a driving protocol that embeds an exact periodic trajectory into the model.

In order to embed the periodic trajectory into the SSH chain, we block together sites $\{2n,2n{+}1 \}$ and use a $d{=}4$, $\chi{=}1$ MPS ansatz. The SSH Hamiltonian in Eq.~(\ref{eq:SSH}) then neatly fits into the form introduced above, with $\hat{\mathcal{H}}_0$ being the $J_o$ term and $\hat{\mathcal{H}}_1$ the $J_e$ term.  It is straightforward to see that  Eq.~(\ref{eq:cond1}) is satisfied. To see that  Eq.~(\ref{eq:cond2}) is satisfied, we note that because the variational parameters are localized to a single site, each term in the sum defining $P_{\mathcal{T}}$ differs from $\ket{\psi(t)}$ on just one site. $\hat{\mathcal{H}}_1$ acting on $\ket{\psi(t)}$ makes it orthogonal to $\ket{\psi(t)}$ on two sites, therefore $\hat{\mathcal{H}}_1\ket{\psi(t)}$ is annihilated by $P_{\mathcal{T}}$.  In this sense, the SSH Hamiltonian for any $J_e$ has the same semiclassical limit, corresponding to the quantum dynamics of the $J_e=0$ model.

Suppose we modify the SSH chain by adding longer range hopping terms of the form:
\begin{eqnarray}
\label{eq:SSH2}
\nonumber \hat{\mathcal{H}} \hspace{-0.1cm}&=& \hspace{-0.2cm}\sum_n\hspace{-0.1cm} J_o \sigma^+_{2n+1}\sigma^-_{2n+2}+J_e \sigma^+_{2n+2}\sigma^-_{2n+3}{+} \Delta \sigma^+_{2n+1}\sigma^-_{2n+4}\\
&+& i\alpha (-1)^{n}\left(\sigma^+_{2n+1}\sigma^-_{2n+3} -\sigma^+_{2n+2}\sigma^-_{2n+4}\right)+ \mathrm{h.c.}
\end{eqnarray}
The additional hopping terms introduced here all satisfy Eqs.~(\ref{eq:cond1})-(\ref{eq:cond2}) and therefore Eq.~(\ref{eq:SSH2}) defines a class of models that share the same semiclassical limit as the SSH chain. This form was chosen so the $\hat{\mathcal{H}}_1$ contributions all take $\ket{\psi(t)}$ to the same state. For this reason the quantum leakage takes a simple form,
\begin{equation}
\Gamma = \frac{\sqrt{N-2}}{T}\int^{T=\frac{\pi}{J_o}}_{t=0} \left|\frac{J_e+\Delta}{2}\sin(2 J_o t)+\alpha\right|\d t.
\end{equation}
When $\Gamma = 0$, the periodic TDVP trajectory becomes an exact trajectory in the full quantum dynamics. By fixing $\alpha = 0$ and $J_e=-\Delta$, we obtain a  family of static Hamiltonians, Eq.~(\ref{eq:SSH2}), that admit exact periodic orbits. However, we can also make $\Gamma$ vanish if we allow the coupling to vary with time, $\alpha(t)=-\frac{1}{2}(J_e+\Delta)\sin(2J_ot)$, as  confirmed in Fig.~\ref{fig:Fig2}(a). The latter Floquet model hosts the same periodic orbit as the static SSH model.  However, unlike the static case, the tower of QMBS eigenstates are not preserved by the Floquet operator, see Fig.~\ref{fig:Fig2}(c). This is reminiscent of Rydberg atoms with a modulated chemical potential~\cite{Bluvstein2021}, where the scarred initial state also has high overlap with only a few Floquet modes~\cite{Hudomal2022Driven}.  

{\em AKLT model.---}Our construction can also embed trajectories that involve entangled states with nontrivial correlations.  As a second example, we consider the AKLT model~\cite{Affleck1987rigorous} -- a paradigmatic model of symmetry protected topological (SPT) order:
\begin{equation}
\hat{\mathcal{H}}_\mathrm{AKLT}=\sum_n \mathbf{S}_n\cdot \mathbf{S}_{n+1}+\frac{1}{3}(\mathbf{S}_n\cdot \mathbf{S}_{n+1})^2,
\end{equation}
where $\mathbf{S}_n$ is a spin-1 operator on lattice site $n$. Recently, there has been much interest in quantum simulations of this model~\cite{Chen2023,Choo2018,Azses2020,Tan2023,smith2022}. For present purposes, it will be important that $\hat{\mathcal{H}}_\mathrm{AKLT}$ contains a tower of QMBS eigenstates, generated by repeatedly applying the $\pi$-momentum spin-raising operator, $Q^+=\sum_n (-1)^n (S^+_n)^2$,  to the ground state \cite{Moudgalya2018, BernevigEnt}. 
The QMBS towers were shown to allow signatures of SPT order, such as the fractionalized boundary excitations, to persist at high energies above the ground state~\cite{Jeyaretnam2021}.

To construct the AKLT periodic orbit, we use the following $\chi{=}2$ initial state,  
\begin{equation}
\label{eq:AKLTstate}
\ket{\psi(0)}=\bigotimes_n \left(\mathbb{1}+(-1)^n(S_n^+)^2/2\right) \ket{\psi^\mathrm{AKLT}_\mathrm{GS}}.
\end{equation}
This state oscillates periodically at a constant entanglement entropy $S_E(t){=}\log(2)$, with the period set by the level spacing in the scarred subspace, $\epsilon{=}4$. 
Note that this choice of the initial state is not unique, e.g., a similar $\chi{=4}$ state was considered in Ref.~\cite{MotrunichTowers}.
 We construct $\hat{\mathcal{H}}_1$ that satisfies Eqs.~(\ref{eq:cond1})-(\ref{eq:cond2}) by noting that for the AKLT ground state, no two neighboring sites can be in the state $\ket{-}$, a property inherited by $\ket{\psi(t)}$. As $P_{\mathcal{T}}$ differs from the MPS state $\ket{\psi(t)}$ on a single site, $\hat{\mathcal{H}}_1$ will satisfy Eq.~(\ref{eq:cond2}) provided it maps at least four neighboring sites to $\ket{-}$.  Therefore, introducing the state $\ket{\chi_{-}}\equiv \ket{-,-,-,-}$, a suitable Hamiltonian will be of the form $\hat{\mathcal{H}}_1= \gamma \sum \ket{\Phi} \bra{\chi_{-}} + \mathrm{h.c.}$, where $\ket{\Phi}$ is an arbitrary state on four sites, which needs to have a finite overlap with $\ket{\psi(t)}$ in order to satisfy Eq.~(\ref{eq:cond1}). These perturbations to the AKLT model differ fundamentally from those that preserve the entire tower of QMBS eigenstates in Ref.~\cite{MotrunichTowers}. Indeed, the QMBS eigenstates of the AKLT model are not contained within the manifold, therefore even perturbations with a perfectly coherent scarred orbit are not required to preserve the eigenstates.  

 Using quantum leakage, we can construct a driven perturbation of the AKLT model with an exact Floquet scarred state.   First, we introduce the local basis vectors, $\ket{\alpha_{\pm}} = (\ket{+}\pm \ket{-})/\sqrt{2}$. Using this basis, we examine the following two-parameter perturbation:
\begin{eqnarray}\label{eq:AKLTpert}
\nonumber \hat{\mathcal{H}}_1 = &\sum_n &  \gamma\ket{\alpha_+,\alpha_-,\alpha_+,\alpha_-}\bra{\chi_{-}} \\
\nonumber &+& \gamma\ket{\alpha_-,\alpha_+,\alpha_-,\alpha_+}\bra{\chi_{-}} \\ 
&+& (-1)^n\Delta\ket{0,+,0,+}\bra{\chi_{-}}+\mathrm{h.c.}
\end{eqnarray}
All of the terms in this $\hat{\mathcal{H}}_1$ map $\ket{\psi(t)}$ to the same state, therefore the leakage takes the form:
\begin{equation}
\Gamma \propto \sqrt{N}\int^{T=\frac{\pi}{2}}_{t=0} \left|\gamma\cos(\epsilon t)-\Delta/2 \right| \d t .
\end{equation}
The leakage can be exactly cancelled by setting $\Delta=2\gamma\cos{(\epsilon t)}$, as confirmed in Fig. \ref{fig:Fig2}(b). The Floquet model once again destroys the underlying tower of QMBS states, as shown in Fig. \ref{fig:Fig2}(d).

 While to the best of our knowledge, there is no general relation between QMBS states and SPT order, it is interesting to note that our periodic scarred trajectory exhibits a constant-in-time AKLT string order parameter~\cite{Kohmoto1992}. This is surprising given that the Floquet model breaks the dihedral $\mathbb{Z}_2 {\times} \mathbb{Z}_2$ symmetry that normally protects the SPT order in the AKLT model~\cite{Pollmann2012}. Thus, our construction can embed a trajectory with quantized SPT order parameter into a non-SPT model. In SM~\cite{SOM}, we show that similar conclusions hold for the cluster model~\cite{ClusterModel}, which exhibits Majorana boundary modes~\cite{Pachos2004,Smacchia2011}.  

{\em Conclusions and discussion.---}We have presented a method for constructing classes of quantum Hamiltonians with equivalent semiclassical dynamics. This construction results in models that possess approximate QMBS associated with a semiclassical trajectory, reminiscent of scars in quantum billiards~\cite{Heller84}. For the choice of Hamiltonians above,  the calculation of the quantum leakage is tractable, allowing to write down new Floquet models with exact QMBSs (see~\cite{SOM} for further examples). The choice of MPS states for defining the manifold $\mathcal{M}$ was due to many QMBS models previously studied in the literature using MPS methods. However, our approach can be extended to other classes of variational wave functions such as bosonic or fermionic Gaussian states \cite{Guaita2019} or projected entangled pair states (PEPS)~\cite{CiracRMP}. 

The approach here complements recent works that construct exact QMBSs using cellular automata~\cite{Iadecola2020,Rozon2022,Rozon2023}. 
In particular, it furnishes a constructive realization of orbit ``steering" 
by Ljubotina~\emph{et al.}~\cite{Ljubotina2022Steering}. In contrast to the latter, our  approach yields exact Floquet QMBSs without the need for variational optimization. Furthermore, our method does not require that the periodic orbit be generated by QMBS, e.g., it could result from other ergodicity-breaking mechanisms, such as integrability or Hilbert space fragmentation \cite{Yoshinaga2022,Hart2022,Balducci2022}.

If the states in $\cal M$ form an overcomplete basis, then a Feynman path integral over the manifold can be constructed~\cite{green2016feynman}. The saddle point equations of the path integral will correspond to the TDVP equations of motion, while additional perturbative corrections eventually reproduce the exact quantum dynamics. In particular, the quadratic corrections to TDVP equations of motion can be related to Lyapunov exponents which characterise the chaotic nature of mixed semiclassical phase space~\cite{Hallam2019,Michailidis2020}. For Hamiltonians which can be decomposed in the manner introduced in this paper, it is possible to write analytic expressions for the Lyapunov exponents \cite{Hallam-prep}.  

In some physical applications, one would wish to ``invert" the above procedure, i.e., given a manifold $\cal M$ and a Hamiltonian $\hat{\mathcal{H}}$, describing some physical system which supports QMBS, one would like to identify a decomposition into $\hat{\mathcal{H}}_0$ and $\hat{\mathcal{H}}_1$, such that Eqs.~(\ref{eq:cond0})-(\ref{eq:cond2}) approximately hold. A notable example is the PXP model~\cite{FendleySachdev, Lesanovsky2012}, which provides an effective description of QMBS in Rydberg atom arrays. In the PXP model, it is not obvious how to perform the decomposition into $\hat{\mathcal{H}}_0$ and $\hat{\mathcal{H}}_1$, although it has been conjectured that a suitable deformation of the model could result in exact QMBS~\cite{Khemani2018,Choi2018,Omiya2022}. In this context, we note that, while Eq.~(\ref{eq:MPScond}) or Eq.~(\ref{eq:MPScond_inf}) are sufficient conditions to satisfy Eq.~(\ref{eq:cond2}), they are not necessary. Hence, it would be interesting to understand if there exist more general yet analytically tractable mechanisms for embedding periodic orbits into larger families of non-integrable quantum Hamiltonians.

\begin{acknowledgments}
{\em Acknowledgments.---}We acknowledge support by EPSRC grant EP/R513258/1 and by the Leverhulme Trust Research Leadership Award RL-2019-015. Statement of compliance with EPSRC policy framework on research data: This publication is theoretical work that does not require supporting research data.
\end{acknowledgments}

\bibliography{bibliography}

\clearpage 
\pagebreak

\onecolumngrid
\begin{center}
\textbf{\large Supplemental Online Material for ``Embedding semiclassical periodic orbits into chaotic many-body Hamiltonians" }\\[5pt]
Andrew Hallam$^{1}$, Jean-Yves Desaules$^{1}$, and Zlatko Papi\'c$^{1}$ \\
{\small \sl School of Physics and Astronomy, University of Leeds, Leeds LS2 9JT, UK}

\vspace{0.1cm}
\begin{quote}
{\small In this Supplementary Material, we derive the conditions for the embedding of a periodic orbit in cases where the variational manifold is spanned by matrix product states. We demonstrate that the perturbed models remain chaotic, and we clarify their distinction from the eigenstate embedding constructions by Shiraishi and Mori~\cite{ShiraishiMori}.  We provide details of the quantum leakage computation for the Su-Schrieffer-Heeger (SSH) model and Affleck-Kennedy-Lieb-Tasaki (AKLT) model discussed in the main text. We illustrate that our approach can be applied to other scarred models, such as the spin-1 XY model from Ref.~\cite{Iadecola2019_2} on hypercubic lattices in arbitrary dimensions,  a model with an emergent kinetic constraint from  Ref.~\cite{Iadecola2019_3}, and a cluster model that supports symmetry-protected topological (SPT) order~\cite{ClusterModel}.   }\\[10pt]
\end{quote}
\end{center}
\setcounter{equation}{0}
\setcounter{figure}{0}
\setcounter{table}{0}
\setcounter{page}{1}
\setcounter{section}{0}
\makeatletter
\renewcommand{\theequation}{S\arabic{equation}}
\renewcommand{\thefigure}{S\arabic{figure}}
\renewcommand{\thesection}{S\Roman{section}}
\renewcommand{\thepage}{\arabic{page}}
\renewcommand{\thetable}{S\arabic{table}}

\vspace{0cm}

\section{Sufficient conditions for a TDVP periodic orbit in the MPS manifold}

In this section we prove the sufficient conditions that the MPS transfer matrix, and thereby perturbation $\hat{\mathcal{H}}_1$, need to obey in order to sustain a periodic orbit in the TDVP manifold. We assume the perturbing Hamiltonian is a sum of tensor products of $2K$-local operators,
\begin{equation}
\hat{\mathcal{H}}_1=\sum_n \hat{O}^{n}_1 \otimes \hat{O}^{n+1}_2 \otimes \cdots \otimes \hat{O}^{n+2K-1}_{2K},
\end{equation}
and denote the MPS transfer matrices as
\begin{equation}
\mathbb{E}_n(\hat{O}_k)=\sum_{s,\bar{s}^\prime}\bar{A}^{s^\prime}_n \hat{O}^n_{k,s,\bar{s}^\prime} A^s_n.
\end{equation}
Working on a finite spin chain with open boundary conditions, we parametrize the MPS tangent space in the following way:
\begin{equation}
\ket{\partial_{A^n}\psi}\rightarrow \sum_n \sum_{\mathbf{\{\sigma_n}\}}(\ldots A^{n-1}B^nA^{n+1} \ldots )\ket{ \ldots \sigma_{n-1}\sigma_n\sigma_{n+1} \ldots}.
\end{equation}
 For the sake of this discussion, $B^n$ is not required to take any particular form. We also introduce the environment of the tensor network contracted from the left and right, respectively,
\begin{equation}
(\mathbb{L}_n|=\prod^{n}_{m=1} \mathbb{E}^0_m , \qquad |\mathbb{R}_n)=\prod^{N}_{m=n} \mathbb{E}^0_m,
\end{equation}
where $\mathbb{E}^0_n$ is a shorthand notation for $\mathbb{E}_n(\mathbb{1})$.
Finally, let us introduce the mixed MPS transfer matrix 
\begin{equation}
\mathbb{E}^{\bar{B}}_n(\hat{O}_k)=\sum_{s,\bar{s}^\prime}\bar{B}^{s^\prime}_n O^n_{k,s,\bar{s}^\prime} A^s_n,
\end{equation}
with $\mathbb{E}^{\bar{B},0}_n$ representing the case $O=\mathbb{1}$. Using these expressions, we can show that $\braket{\partial_{A^n}\psi|\hat{\mathcal{H}}_1|\psi}=0$ provided that 
\begin{equation}
 \prod^{K}_{k=1} \mathbb{E}_{n+k-1}(\hat{O}_k)=0, \quad \prod^{2K}_{k=K+1}  \mathbb{E}_{n+k-1}(\hat{O}_k)=0.
\end{equation}

In order to demonstrate this let us analyze the various possible cases in turn. Firstly, when the Hamiltonian is one of $2K$ sites that do not overlap with the tangent vector MPS, we find contributions like
\begin{equation}\label{eq:term1}
(\mathbb{L}_{n-1}|\left(\prod^{2K}_{k=1}\mathbb{E}_{n+k-1}(\hat{O}_k)\right)\left( \prod^{j^\prime-1}_{j=n+2K}\mathbb{E}^{0}_n\right)\mathbb{E}^{\bar{B},0}_{j^\prime}|\mathbb{R}_{j^\prime+1}),
\end{equation}
which is clearly zero if $\prod^{K}_{k=1} \mathbb{E}_{n+k-1}(\hat{O}_k)=0$ or $\mathbb{E}_{n+k-1}(\hat{O}_k)=0$. When the tangent vector is located to the left of the Hamiltonian, the contributions will similarly be zero. When the tangent vector overlaps with the Hamiltonian, we will find contributions like  
\begin{equation}\label{eq:term2}
(\mathbb{L}_{n-1}|\left(\prod^{k^\prime-1}_{k=1}\mathbb{E}_{n+k-1}(\hat{O}_k)\right)\mathbb{E}^{\hat{B}}_{n+k^\prime-1} (\hat{O}_{k^\prime})\left( \prod^{2K}_{k=k^\prime+1} \mathbb{E}_{n+k-1}(\hat{O}_k)\right)|\mathbb{R}_{n+2K}).
\end{equation}
Since the tangent vector will be in either the first half or the second half of the Hamiltonian, either $\prod^{K}_{k=1} \mathbb{E}_{n+k-1}(\hat{O}_k)=0$ or $\mathbb{E}_{n+k-1}(\hat{O}_k)=0$ will force this contribution to be zero. These contributions are zero regardless of the particular form of $(\mathbb{L}_{n}|$ and $|\mathbb{R}_{n})$, for this reason they also apply naturally to MPS with periodic boundary conditions. If we work in the thermodynamic limit, with a translation invariant, injective MPS tensor $A^s$, the MPS transfer matrix will have a dominant left and right eigenvectors with eigenvalue $\lambda=1$,
\begin{equation}
(\mathbb{L}|\mathbb{E}_0=(\mathbb{L}| , \qquad \mathbb{E}_0|\mathbb{R})=|\mathbb{R}).
\end{equation}
The equations equivalent to Eqs.~(\ref{eq:term1})-(\ref{eq:term2}) now begin and end with the same eigenvectors everywhere,  $(\mathbb{L}|$ and $|\mathbb{R})$. Therefore, the conditions for $\braket{\partial_{A^n}\psi|\hat{\mathcal{H}}_1|\psi}=0$ can be weakened to 
\begin{equation}
(\mathbb{L}|\prod^{K}_{k=1} \mathbb{E}_{n+k-1}(\hat{O}_k)=0, \quad \prod^{2K}_{k=K+1} \mathbb{E}_{n+k-1}(\hat{O}_k)|\mathbb{R})=0,
\end{equation}
as stated in the main text. This condition can be easily generalized to an  $n$-site translation invariant MPS in the thermodynamic limit. 

\subsection{Example}

To clarify how we can use the construction above to find a valid perturbing Hamiltonian, we give an example for the AKLT model, which was considered in the main text. 
%
The spin-1 AKLT model is defined by the Hamiltonian:
\begin{equation}
\hat{\mathcal{H}}_\mathrm{AKLT}=\sum_i \mathbf{S}_i\cdot \mathbf{S}_{i+1}+\frac{1}{3}(\mathbf{S}_i\cdot \mathbf{S}_{i+1})^2.
\end{equation}
The AKLT ground state is defined by the three MPS tensors
\begin{equation}\label{eq:AKLTMPS}
A^+=\sqrt{\frac{2}{3}}\begin{pmatrix}
0 & 1\\
0 & 0
\end{pmatrix}, \quad 
A^0=\sqrt{\frac{1}{3}}\begin{pmatrix}
-1 & 0\\
0 & 1
\end{pmatrix}, \quad
A^-=-\sqrt{\frac{2}{3}}\begin{pmatrix}
0 & 0\\
1 & 0
\end{pmatrix}.
\end{equation}

For this wavefunction $(\mathbb{L}|=(1,0,0,1)$ and $|\mathbb{R})=(1,0,0,1)^T/2$. Suppose we want to build a perturbation from the operators $(S^+_n)^2$ and $(S^-_n)^2$. In this case, we find that
\begin{equation}
\mathbb{E}_n((S^+_n)^2)=-\frac{4}{3}\begin{pmatrix}
0 & 0 & 0 & 0\\
0 & 0 & 0 & 0\\
0 & 1 & 0 & 0\\
0 & 0 & 0 & 0
\end{pmatrix}
\quad  \text{and} \quad 
\mathbb{E}_n((S^-_n)^2)=-\frac{4}{3}\begin{pmatrix}
0 & 0 & 0 & 0\\
0 & 0 & 1 & 0\\
0 & 0 & 0 & 0\\
0 & 0 & 0 & 0
\end{pmatrix}.
\end{equation}
Since $\mathbb{E}_n((S^+_n)^2)|\mathbb{R})=\mathbb{E}_n((S^-_n)^2)|\mathbb{R})=(\mathbb{L}|\mathbb{E}_n((S^+_n)^2)=(\mathbb{L}|\mathbb{E}_n((S^-_n)^2)=0$, we can construct a two-site Hamiltonian of the form
\begin{equation}
    \mathcal{H}_1=\sum_n (S^+_n)^2(S^+_{n+1})^2+(S^-_n)^2(S^-_{n+1})^2.
\end{equation}
Provided we work in the thermodynamic limit, this is a valid perturbation.

\section{Relation to Shiraishi-Mori eigenstate embedding}

Here we discuss the relation between our approach and that of Shiraishi-Mori~\cite{ShiraishiMori}, which embeds target eigenstates in a generally 
thermalizing spectrum. The time-dependent variational principle can be reframed as a modified Schr\"odinger equation where the Hamiltonian is dressed with time-dependent projectors. Suppose we introduce a  projector which projects onto either the state $P_0(t)=\ket{\psi(t)}\bra{\psi(t)}$, or onto the TDVP tangent space of this state, $P_\mathcal{T}(t)$. These two projectors can be combined to define a single  $P(t)=P_0(t)+P_\mathcal{T}(t)$ with complement  $Q(t)=\mathbb{1}-P(t)$. From this perspective, we can rewrite the Hamiltonian as
\begin{equation}
\mathcal{H}=P(t)\mathcal{H}P(t)+P(t)\mathcal{H}Q(t)+Q(t)\mathcal{H}P(t)+Q(t)\mathcal{H}Q(t).
\end{equation}
The first term of this decomposition, $P(t)\mathcal{H}P(t)$ corresponds to the TDVP equations. If we group together the remaining terms of $\mathcal{H}^\prime(t)=P(t)\mathcal{H}Q(t)+Q(t)\mathcal{H}P(t)+Q(t)\mathcal{H}Q(t)$, this superficially resembles the Shiraishi-Mori construction of eigenstate embedding. However, upon further examination it is clear that this decomposition it not related. Firstly, the time-dependent projectors $P(t)$ do not annihilate the relevant state, $\ket{\psi(t)}$. Secondly $[P(t),\mathcal{H}^\prime(t)]\neq 0$. For this reason, the approach discussed in this work should not be viewed as a straightforward generalization of Shiraishi-Mori to a time-dependent setting.

\section{SSH model: leakage calculation}

In the main text we considered a modified SSH model on a $N$-site chain with open boundary conditions (OBC). We can divide the model up into $\hat{\mathcal{H}}_0$ and $\hat{\mathcal{H}}_1$ terms, where $\hat{\mathcal{H}}_0$ takes the form
\begin{equation}
\begin{split}
\hat{\mathcal{H}}_0=&\sum^{N/2-1}_{n=0} J_o \sigma^+_{2n+1}\sigma^-_{2n+2}+ \mathrm{h.c.}
\end{split}
\end{equation}
For this Hamiltonian, the initial state $\ket{\psi(0)}=\bigotimes^{n=N/4-1}_{n=0} \ket{1,0,0,1}$ oscillates periodically, taking the form 
\begin{equation}
\ket{\psi(t)}=\bigotimes^{N/4-1}_{n=0} \ket{\phi_1(t)}\otimes\ket{\phi_2(t)}=\bigotimes_{n} \left( \cos(J_ot)\ket{10}-i\sin(J_ot)\ket{01}\right) \otimes \left(\cos(J_ot)\ket{01}-i\sin(J_ot)\ket{10} \right) ,
\end{equation}
where $\ket{\phi_1(t)}$ is the same for every pair of sites $\{4n+1,4n+2\}$ and, similarly, $\ket{\phi_2(t)}$ is on sites $\{4n+3,4n+4\}$.
For this reason, it is appropriate to block together sites $\{4n+1,4n+2\}$ and  $\{4n+3,4n+4\}$. This $d=4$, $\chi=1$ MPS ansatz defines the variational manifold for which we calculate the TDVP equations and quantum leakage.  

\begin{figure}[tb]
\centering
    \includegraphics[width=1.0\linewidth]{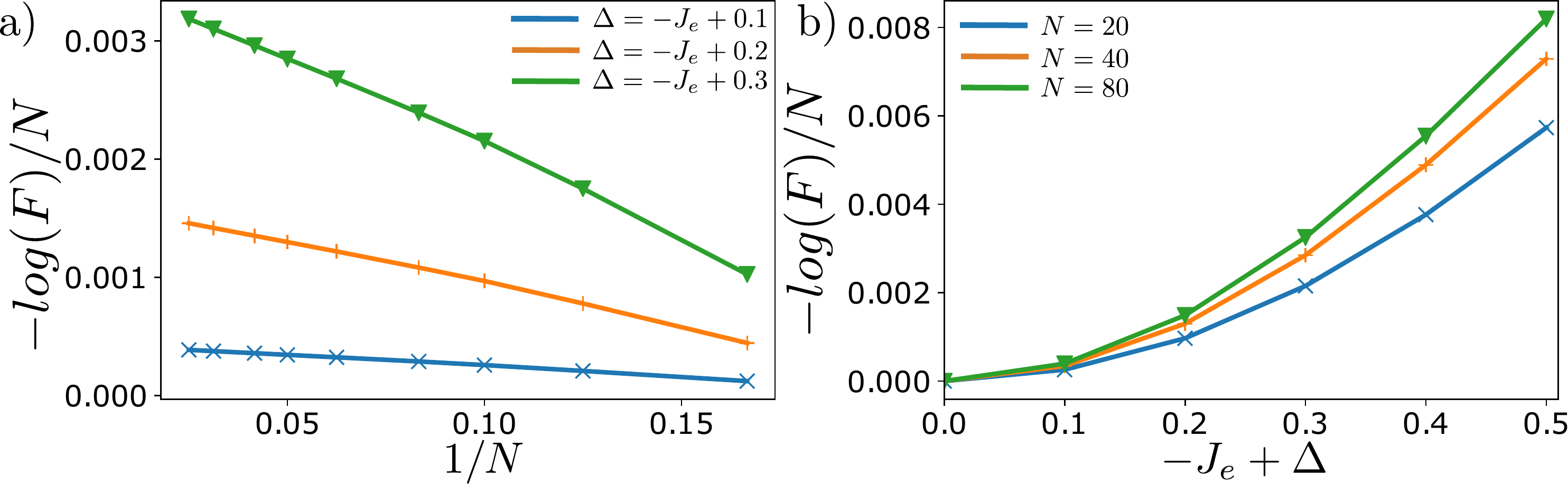}
      \caption{ $-\log(F)/N$ for the SSH model with $J_o=1$, $J_e=2/3$, $\alpha=0$ and varying $\Delta$. a) shows $-\log(F)/N$ against $1/N$ for various $\Delta$, it approaches a value far from $\log(2)$ in the thermodynamic limit, indicating strong revivals. b) shows $-\log(F)/N$ for various $N$ as a function of $\Delta$, it appears to grow approximately quadratically.  }
         \label{fig:Fig_SSH_dF}
\end{figure}

Given this, we can choose the $\hat{\mathcal{H}}_1$ Hamiltonian to be,
\begin{equation}
\begin{split}
\hat{\mathcal{H}}_1=&\sum^{N/2-2}_{n=0} J_e \sigma^+_{2n+2}\sigma^-_{2n+3}+ \Delta \sigma^+_{2n+1}\sigma^-_{2n+4}+ i\alpha (-1)^{n}\left(\sigma^+_{2n+2}\sigma^-_{2n+4} -\sigma^+_{2n+1}\sigma^-_{2n+3}\right)+\mathrm{h.c.} \\
\end{split}
\end{equation}
We find that the instantaneous quantum leakage is
\begin{equation}
\Gamma(t)=||\hat{\mathcal{H}}_1\ket{\psi(t)}||.
\end{equation}
We can evaluate this by calculating the action one term of $\hat{\mathcal{H}}_1$ on $\ket{\psi(t)}$. A single term of $\hat{\mathcal{H}}_1$ acting on $\ket{\phi_1(t)}\otimes\ket{\phi_2(t)}$ in $\ket{\psi(t)}$ is
\begin{eqnarray}
\left(\bigotimes \ket{\phi_1(t)}\otimes\ket{\phi_2(t)}\right) \otimes \left( -i\left((J_e+\Delta)\cos(J_ot)\sin(J_ot)+\alpha\right)\right)\left(\ket{0011}+\ket{1100}\right) \left(\bigotimes \ket{\phi_1(t)}\otimes\ket{\phi_2(t)}\right). 
\end{eqnarray}
A similar result is obtained when $\hat{\mathcal{H}}_1$ acts on $\ket{\phi_2(t)}\otimes\ket{\phi_1(t)}$ instead.
So in total there are $N/2-1$ terms in $\hat{\mathcal{H}}_1$, all of which have zero overlap with one another and contribute similarly. For this reason, we find that the total instantaneous quantum leakage is
\begin{eqnarray}
    \nonumber \Gamma(t) &=&  \| \sum_{n=0}^{N/2-2} \ket{\psi_{L,2n}(t)}\otimes\left(\ket{0011}+\ket{1100}\right)\otimes \ket{\psi_{R,2n+5}(t)}\| \times \left|(J_e+\Delta)\cos(J_ot)\sin(J_ot)+\alpha\right|
    \\ \nonumber && = \sqrt{N/2-1} \times\sqrt{2}\times\left|\frac{1}{2}(J_e+\Delta)\sin(2J_ot)+\alpha\right|
        \\ && = \sqrt{N-2}\times\left|\frac{1}{2}(J_e+\Delta)\sin(2J_ot)+\alpha\right|,
\end{eqnarray}
where $\ket{\psi_{L,2n}(t)}$
and $\ket{\psi_{R,2n+5}(t)}$ denote the parts of $\ket{\psi(t)}$ on which the term of $\hat{\mathcal{H}}_1$ is not acting, and which is simply composed of a tensor product of $\ket{\phi_1 (t)}$ and $\ket{\phi_2 (t)}$.
When the leakage does not exactly cancel, quantum dynamics will only approximately be periodic. We characterize such  approximate periodic dynamics using the maximum fidelity revival, $F=\max_{t \in [t_1,t_2]} |\langle \psi(t)|\psi(t=0)\rangle|^2$, over some moderate time interval $[t_1,t_2]$ that exceeds an initial relaxation (given by the energy scales of the microscopic Hamiltonian). For a generic quantum trajectory, the fidelity density $-\log(F)/N$ is expected to approach $\log(d)$ in the thermodynamic limit, where $d$ is the local Hilbert space dimension on each site. In Fig.~(\ref{fig:Fig_SSH_dF})(a), we see that the fidelity density remains small for various $\Delta$ as the system size increases. Moreover, in Fig.~(\ref{fig:Fig_SSH_dF})(b), we see $-\log(F)/N$ has an approximately quadratic dependence on $-J_e+\Delta$.  

We note that the same results hold for periodic boundary conditions as long as $N$ is a multiple of 4. The only difference is that he prefactor in the leakage becomes $\sqrt{N}$ instead of $\sqrt{N-2}$.

As a side note, we can simply generalize the driven model one with long-range interactions due to the fact that the wavefunction has zero correlation length. For example, any perturbation of the form
\begin{equation}
\begin{split}
\hat{\mathcal{H}}_1=&\sum^{N/2-2}_{n=0}\sum_{m=0} f(m)( J_e \sigma^+_{2n+2}\sigma^-_{2n+3+4m}+ \Delta \sigma^+_{2n+1}\sigma^-_{2n+4+4m}+ i\alpha (-1)^{n}\left(\sigma^+_{2n+2}\sigma^-_{2n+4+4m} -\sigma^+_{2n+1}\sigma^-_{2n+3+4m}\right)+\mathrm{h.c.})
\end{split}
\end{equation}
would be valid. Here $f(m)$ can be arbitrary function,  but $\sum^{m=\infty}_{m=0}f(m)$ should converge for the model to possess a well defined thermodynamic limit.  This model will also feature exact quantum revivals provided $\frac{1}{2}(J_e+\Delta)\sin(2J_ot)+\alpha=0$.

\subsection{Nearest-neighbor coupling}

\begin{figure}[tb]
\centering
    \includegraphics[width=0.5\linewidth]{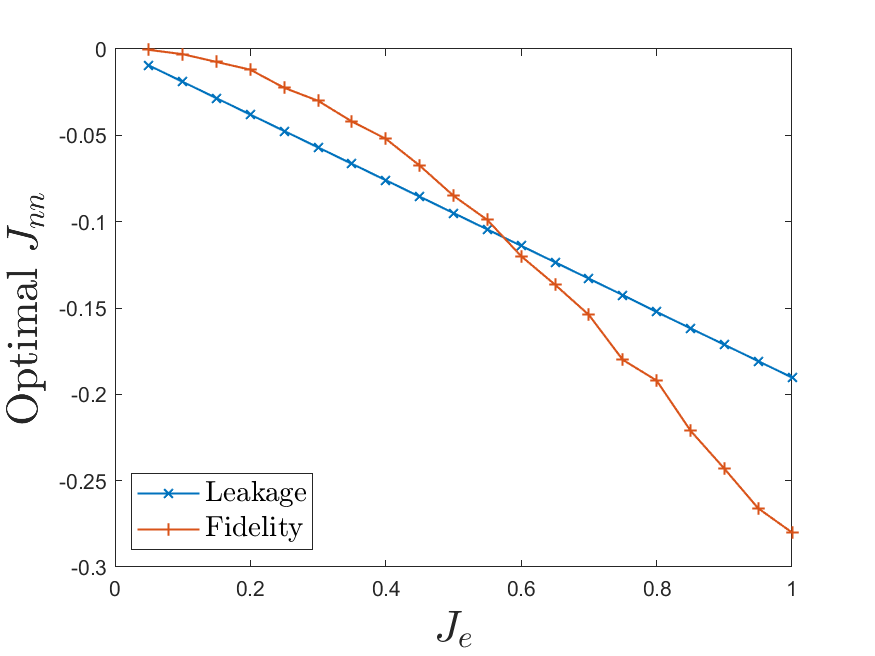}
      \caption{The optimal value of $J_{nn}$ that minimizes quantum leakage, found using Eq.~(\ref{eq:Jnn_leakage}), with $J_o=1$, $N=50$ (blue). This can be compared with the maximum fidelity revival found numerically using $\chi=64$ TDVP with a timestep of $\delta t=0.025$ (orange). }
         \label{fig:Fig_SSH_Jnn}
\end{figure}

If, instead, $\hat{\mathcal{H}}_1$ contains next-next-nearest neighbor term of the form
\begin{equation}
\begin{split}
\hat{\mathcal{H}}_1=&\sum^{N/2-2}_{n=0} J_e \sigma^+_{2n+2}\sigma^-_{2n+3}+ \Delta \sigma^+_{2n+1}\sigma^-_{2n+4}+\sum^{N/2-3}_{n=0}J_{nn}\sigma^+_{2n+2}\sigma^-_{2n+5}+h.c., \\
\end{split}
\end{equation}
we can once again perform the quantum leakage calculation. We find one contribution equivalent to the calculation in the previous section, with an additional term due to $J_{nn}$ acting on next-nearest neighboring sites in the $d=4$ ansatz. Taking into account these terms, we find that the instantaneous quantum leakage is
\begin{equation}
\label{eq:Jnn_leakage}
\Gamma(t)=\sqrt{N/2-1}\sqrt{1/2(J_e+J_{nn})^2\sin(2J_ot)^2+J_{nn}^2\frac{N/2-2}{N/2-1}\left(\sin(J_o t)^4+\cos(J_o t)^4\right)}.
    \end{equation}
As the two terms under the square root are both positive definite, it is clear that there is no choice of $J_{nn}$ which will lead to $\Gamma(t)=0$ for finite $J_e$. Nevertheless, increasing $J_{nn}$ can still reduce the leakage. Fig.~\ref{fig:Fig_SSH_Jnn} shows the optimal choice of $J_{nn}$ that minimizes the leakage. We see that the optimal $J_{nn}$ increases linearly with $J_e$. This is in reasonably good agreement with the optimal choice of $J_{nn}$ found by maximizing the fidelity revival for the exact quantum dynamics.

We note that, in addition to the nearest-neighbor coupling considered above, Ref.~\cite{Zhang2022} also found that a driven perturbation of the form $\propto \sum_n \sin(\Omega t)\sigma^z_n$ can enhance the periodic revivals due to scarring. Unfortunately, it is not possible to treat such perturbations  using the method introduced here, as any single-site operator commutes with the tangent space projector, therefore it is not of the appropriate form required for $\hat{\mathcal{H}}_1$.

\section{Spin-1 XY model}

Another model that  has been rigorously shown to exhibit quantum many-body scars is the spin-1 XY chain~\cite{Iadecola2019_2}, 
\begin{equation}
\hat{\mathcal{H}}_0=\sum_n J(S^x_nS^x_{n+1}+S^y_nS^y_{n+1})+hS^z_n+D(S^z_n)^2.
\end{equation}
Here, we will once again work with open-boundary conditions (OBC). This model features a tower of scarred eigenstates generated by repeatedly applying the operator
\begin{equation}
Q^+ =\sum_n (-1)^n (S^+_n)^2
\end{equation}
on the state $\ket{0}=\otimes_n\ket{m_n=-1}$. The $m$-th eigenstate in this tower has an energy $E_m=h(2m-N)+ND$. Due to the even level spacing of this tower, any state written as a superposition of these eigenstates will oscillate periodically. One particularly nice example is the initial state
\begin{equation}
\ket{\psi(t=0)}=\bigotimes_n \frac{1}{\sqrt{2}}(\ket{m_n=-1}+(-1)^n\ket{m_n=1}),
\end{equation}
for which $\ket{\psi(t)}$ remains an unentangled, product state throughout its evolution. Therefore its behavior is captured entirely by $d=2$, $\chi=1$ MPS. 

 We can choose 
  \begin{equation}
     \hat{\mathcal{H}}_1=\gamma\sum_n (S^x_n S^x_{n+1}-S^y_n S^y_{n+1})+\sum_n (-1)^n\Delta(S^x_n S^y_{n+1}-S^y_n S^x_{n+1})
 \end{equation}
 to satisfy the conditions in Eq.~(4) and Eq.~(5) in the main text. $\hat{\mathcal{H}}_1$ will clearly satisfy Eq.~(4) but Eq.~(5) is more subtle. We can see Eq.~(5) is satisfied by noting that every site in every state in the TDVP ansatz is some superposition of $m_i=-1$ or $m_i=1$ and $\hat{\mathcal{H}}_1$ will flip two sites of the system to be $m_i=0$. The tangent space projector is equal to  $\ket{\psi(t)}$ on all but one site, but $\hat{\mathcal{H}}_1$ will make the state orthogonal to $\ket{\psi(t)}$ on two sites, therefore  Eq.~(5) of the main text will be satisfied.

\begin{figure}[tbh]
\centering
    \includegraphics[width=0.45\linewidth]
    {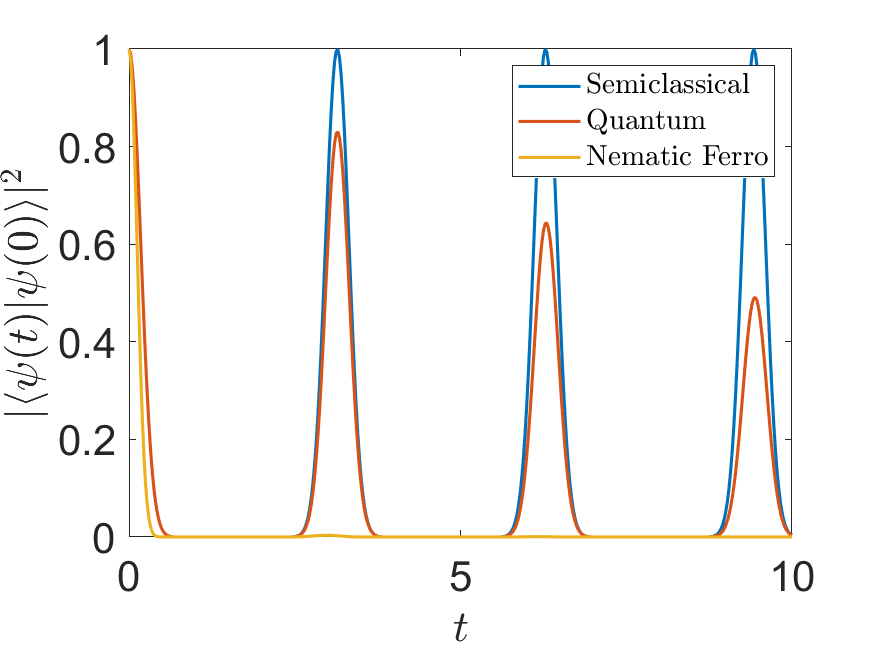}
    \includegraphics[width=0.45\linewidth]
    {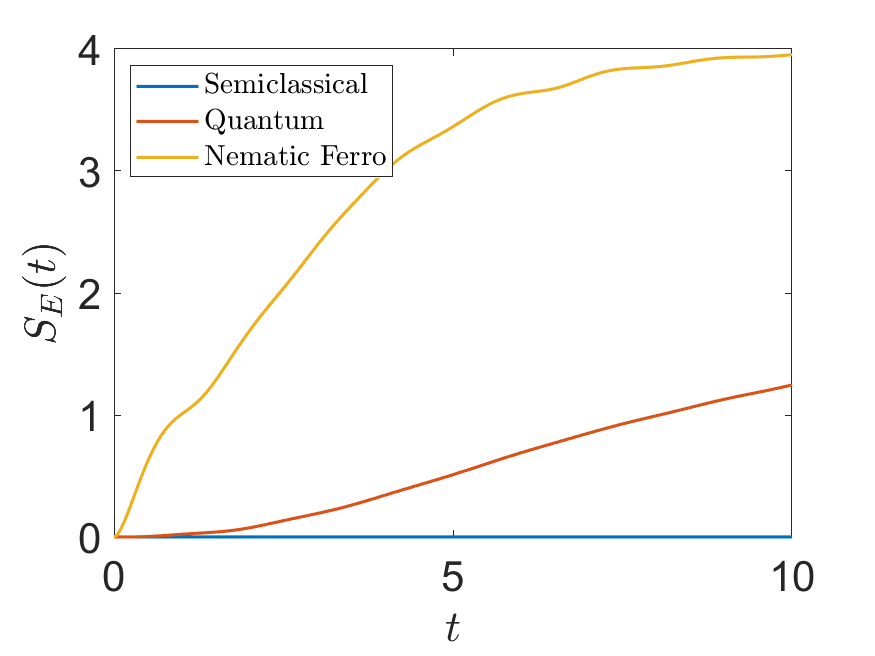}
     \caption{Left: Fidelity revivals for the modified spin-1 XY model with $J=1$, $h=1$, $D=0.5$, $\gamma=0.1$, $\Delta=0$, $N=16$. Perfect fidelity revivals are seen for the semiclassical TDVP ansatz ($\chi=1$) compared to the full quantum evolution ($\chi=100$). No revivals are seem for the generic state, ``Nematic Ferro". Right: Entanglement entropy growth for the same model. The full dynamics still exhibits slow entropy growth compared to a generic state. }
         \label{fig:XYpert}
\end{figure}

In Fig.~\ref{fig:XYpert}(a) we compute the fidelity revivals for the spin-1 XY model after introducing a perturbation with $\gamma=0.1$, $\Delta=0.0$,  which destroys the exact scarring structure. For bond dimension $\chi=1$ we have a product state ansatz and the revivals remain exact but the full quantum dynamics exhibits only imperfect revivals. Fig.~\ref{fig:XYpert}(b) shows the entropy growth. The ansatz has exactly zero entropy for all times but the full dynamics still exhibits slow entropy growth compared to a generic initial state, the nematic ferromagnetic state $\ket{\psi(t=0)}=\bigotimes_i \frac{1}{\sqrt{2}}(\ket{m_i=-1}+\ket{m_i=1})
$.

\begin{figure*}
\centering
    \includegraphics[width=0.32\linewidth]{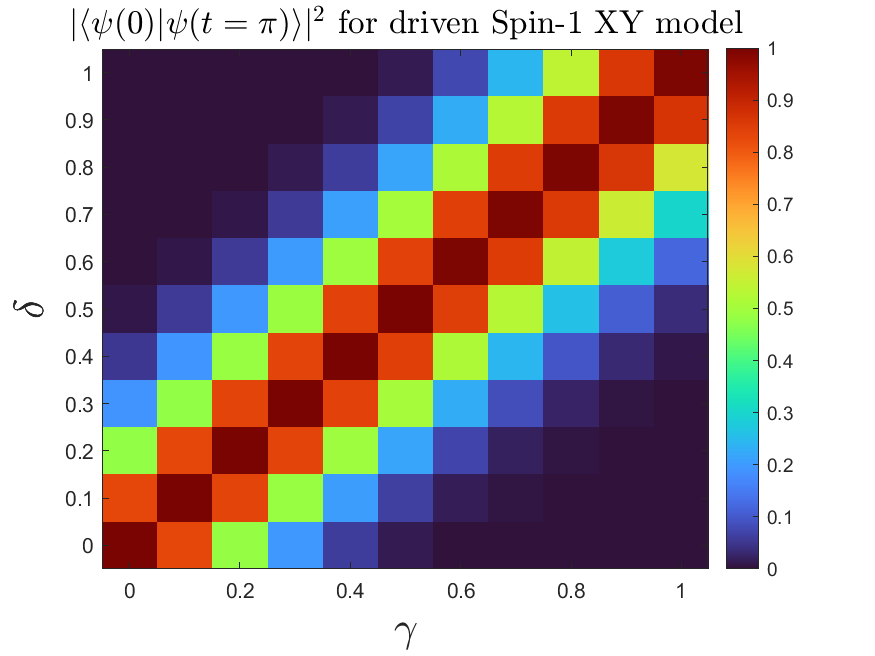}
         \label{fig:XYrevival}
    \includegraphics[width=0.32\linewidth]{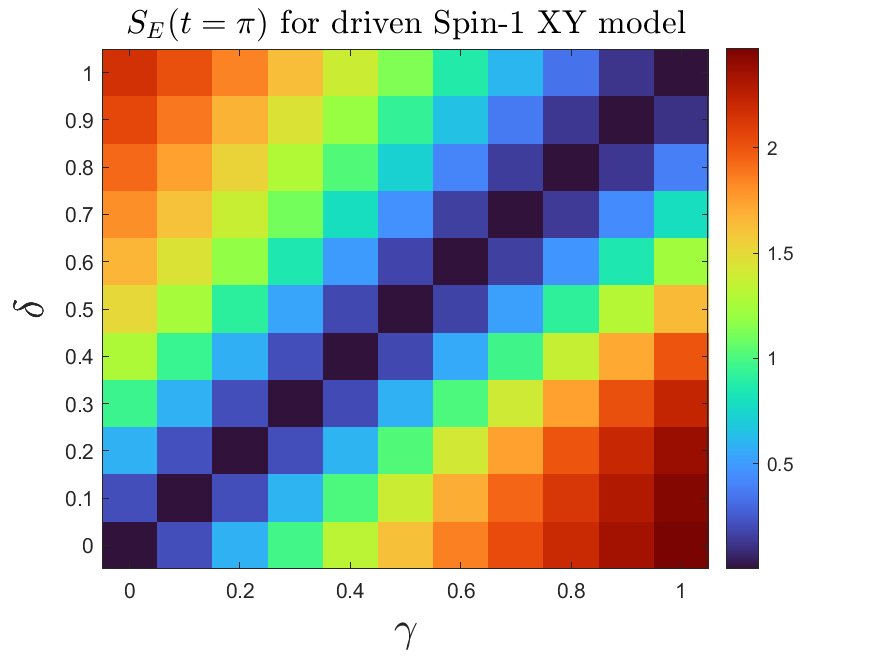}
    \includegraphics[width=0.32\linewidth]
    {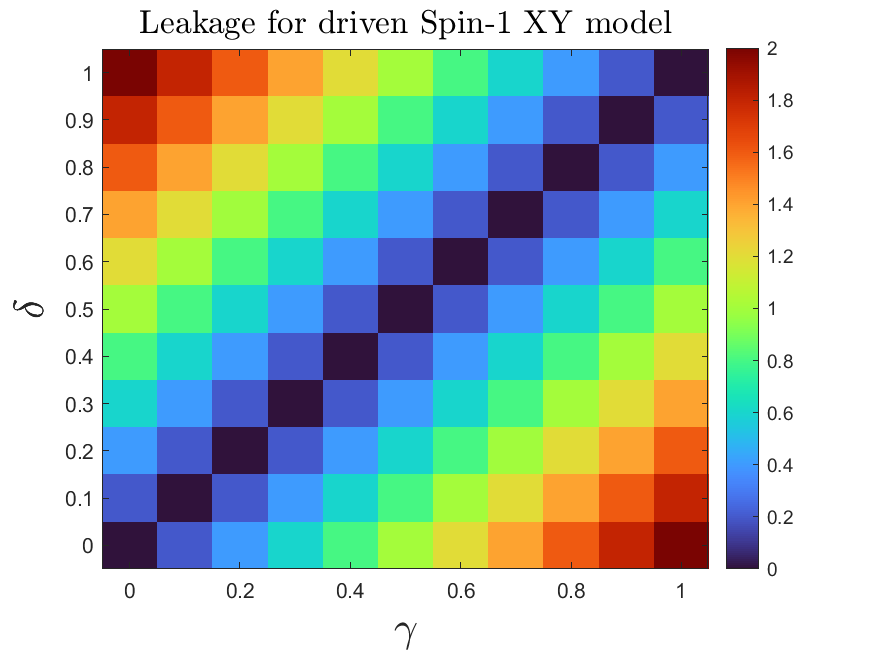}
      \caption{Behavior of the driven spin-1 XY model for fixed  $J=1$, $h=1$, $D=0.5$ and $L=16$ while varying $\delta$ and $\gamma$. Left: Fidelity revival at $t=\pi$. Middle: Entropy at $t=\pi$. Right: Integrated leakage rescaled by the square root of the system size $\Gamma/\sqrt{N}$.}
         \label{fig:XYdriven}
\end{figure*}

 We find that the leakage in this case is 
\begin{equation}
\Gamma = \frac{\sqrt{N-1}}{T}\int^T_0 |\gamma\sin(2ht)-\Delta| \; \d t.
\end{equation}
Therefore, by choosing $\Delta=\gamma \sin(2ht)$, we can once again construct Floquet scars through exact cancellation of the quantum leakage. For this reason, we choose $\Delta=\delta \sin(2ht)$ and vary $\gamma$ and $\delta$.  In Fig.~\ref{fig:XYdriven} we show the behavior of the driven spin-1 XY model for various $\gamma$ and $\delta$. When $\gamma=\delta$ we see perfect fidelity revivals despite it being a strongly driven system. We see that the entropy growth is suppressed to near zero in Fig. \ref{fig:XYdriven}(b). In Fig.~\ref{fig:XYdriven}(c) we show the integrated leakage (divided by system size), which closely predicts the behavior of the entropy and fidelity. 

Since the oscillating state is a product state with zero correlation length it can also be generalized to long-range interactions straightforwardly. For a perturbing Hamiltonian of the form 
  \begin{equation}
     \hat{\mathcal{H}}_1=\gamma\sum_n\sum_{m=0} f(m)( (S^x_n S^x_{n+1+2m}-S^y_n S^y_{n+1+2m})+\sum_n (-1)^n\Delta(S^x_n S^y_{n+1+2m}-S^y_n S^x_{n+1+2m})),
 \end{equation}
 with an arbitrary function $f(m)$, choosing $\Delta=\gamma \sin(2ht)$ still leads to exact quantum leakage cancellation and perfect Floquet scarring.

\subsection{Higher dimensional Spin-1 XY model}

Scarring in the spin-1 XY model can be generalized straightforwardly to hypercubic lattices in any dimension. The Hamiltonian is generalized so the XY interaction acts on neighbouring sites of the hypercubic lattice:
\begin{equation}
\hat{\mathcal{H}}_0=\sum_{\langle n,m\rangle} J(S^x_nS^x_{m}+S^y_nS^y_{m})+hS^z_n+D(S^z_n)^2.
\end{equation}
On this hypercubic lattice the initial state is now
\begin{equation}
\ket{\psi(t=0)}=\bigotimes_n \frac{1}{\sqrt{2}}(\ket{m_n=-1}+\omega(n)\ket{m_n=1}),
\end{equation}
where $\omega(n)=\pm 1$ on even/odd sublattices. We can generalize the driven model to hypercubic lattices similarly:
  \begin{equation}
     \hat{\mathcal{H}}_1=\sum_{\langle n,m\rangle} \gamma(S^x_n S^x_{m}-S^y_n S^y_{m})+ \omega(n)\Delta(t)(S^x_n S^y_{m}-S^y_n S^x_{m}).
 \end{equation}
Choosing $\Delta=\gamma \sin(2ht)$ will once again result in perfect coherent oscillations for this initial state.

\section{Computation of leakage in the AKLT model}

As an example of a model whose periodic orbit features entangled states, in the main text we considered the spin-1 AKLT model, with the ground state MPS matrices given in Eq.~(\ref{eq:AKLTMPS}).  
%
%
In the case of open boundary conditions, the AKLT ground state is four-fold degenerate. The tower of eigenstates are only generated by acting on one of these ground-states, chosen by applying the boundary vectors $v_L=(1,0)$ and $v_R=(0,1)$ to the first and last sites of the system. We can find a bond-dimension two initial state
\begin{equation}
\label{eq:aklt_init}
\ket{\psi(t=0)}=\bigotimes_j (\mathbb{1}+z(-1)^j(S_j^+)^2)\ket{\psi_\mathrm{GS}}
\end{equation}
which has overlap with only the scarred AKLT subspace. In fact, this state is the natural equivalent of the spin-1 XY initial state considered above.  The dynamics from this initial state is remarkably simple:
\begin{equation}
\ket{\psi(t)}=\bigotimes_j (\mathbb{1}+z(-1)^j e^{i \epsilon t}(S_j^+)^2) \ket{\psi_\mathrm{GS}}.
\end{equation}
The state periodically oscillates, with the period set by the level spacing in the scarred subspace, $\epsilon=4$. In terms of the MPS tensors this looks like:
\begin{equation}
A_j^+(t)=\sqrt{\frac{2}{3}}\begin{pmatrix}
0 & 1\\
2(-1)^j z e^{i\epsilon t} & 0
\end{pmatrix}, \quad
A_j^0(t)=\sqrt{\frac{1}{3}}\begin{pmatrix}
-1 & 0\\
0 & 1
\end{pmatrix}, \quad
A_j^-(t)=-\sqrt{\frac{2}{3}}\begin{pmatrix}
0 & 0\\
1 & 0
\end{pmatrix}.
\end{equation}

We can now focus on the case of an infinite chain. For that, we need to compute the two-site transfer matrix for this MPS tensor which is
\begin{equation}
\frac{1}{9}\begin{pmatrix}
    5+16|z|^2 & 0 & 0 & 4 \\
    0 & 1-16|z|^2 & 0 & 0 \\
    0 & 0 & 1-16|z|^2 & 0 \\
    4(4|z|^2+1) & 0 & 0 & 5+16|z|^2\\
\end{pmatrix},
\end{equation}
with eigenvalues
\begin{equation}
    \left\{\frac{1}{9}(16|z|^2+4\sqrt{4|z|^2+1}+5),\frac{1}{9}(16|z|^2-4\sqrt{4|z|^2+1}+5),\frac{1}{9}(1-16|z|^2),\frac{1}{9}(1-16|z|^2)\right\}.
\end{equation} 
We therefore introduce a factor
\begin{eqnarray}
 \mathcal{N}(z)= \left( \frac{1}{9}(16|z|^2+4\sqrt{4|z|^2+1}+5) \right)^{N/2}=\left(\frac{1}{3}(2\sqrt{4|z|^2+1}+1)\right)^N=n_z^N,   
\end{eqnarray}
which appropriately normalizes the state in the thermodynamic limit. Putting the MPS tensor into left canonical form, the dominant left/right eigenvectors become $(\mathbb{L}|=\{1, 0, 0, 1\}$, $|\mathbb{R})=1/2\{1, 0, 0, 1\}^T$, showing the state will evolve at constant entanglement entropy $S_E(t)=\log(2)$.  

If we perform a local change of basis on every site, $\ket{\alpha_{\pm}}=\frac{1}{\sqrt{2}}(\ket{+}\pm \ket{-})$, the time evolved state can be written as
\begin{equation}
A_j^{\alpha_+}(t)=\sqrt{\frac{1}{3}}\begin{pmatrix}
0 & 1\\
-1+2(-1)^j z e^{i\epsilon t} & 0
\end{pmatrix}, \quad
A_j^0(t)=\sqrt{\frac{1}{3}}\begin{pmatrix}
-1 & 0\\
0 & 1
\end{pmatrix}, \quad
A_j^{\alpha_-}(t)=\sqrt{\frac{1}{3}}\begin{pmatrix}
0 & 1\\
1+2(-1)^j z e^{i\epsilon t} & 0
\end{pmatrix}.
\end{equation}
By making the choice $z=1/2$, the bottom left element of $A^{\alpha_+}$ and $A^{\alpha_-}$ can be rewritten respectively as $-2e^{i\epsilon t/2}\cos{(\epsilon t/2)}$ and $-2ie^{i\epsilon t/2}\sin{(\epsilon t/2)}$ for odd sites. For even sites they are $2ie^{i\epsilon t/2}\sin{(\epsilon t/2)}$ and  $2e^{i\epsilon t/2}\cos{(\epsilon t/2)}$ respectively. For $z=1/2$, $n_z=n_{1/2}=\frac{1}{3}(1+2\sqrt{2})$. Using this basis we can study the behavior of the following two parameter perturbation:
\begin{equation}
\hat{\mathcal{H}}_1=\sum_i \gamma(\ket{\alpha_+,\alpha_-,\alpha_+,\alpha_-}\bra{-,-,-,-}+\ket{\alpha_-,\alpha_+,\alpha_-,\alpha_+}\bra{-,-,-,-})+(-1)^i\Delta\ket{0,+,0,+}\bra{-,-,-,-}+h.c.
\end{equation}

Calculating the leakage is more complicated in this case due to the non-zero overlap between $\hat{\mathcal{H}}_1$ acting on different sites. Let us write $\hat{\mathcal{H}}_1=\sum_i h_i$. We first note that on the four sites where $h_i$ modifies the MPS, the only nonzero basis vector is $\ket{-}$ and the product of the four MPS tensors is proportional to the identity, $h^{i,i+1,i+2,i+3}_i A^i A^{i+1}A^{i+2}A^{i+3} \propto \mathbb{1}_2$. Then we note that, $\braket{\psi(t)|h_i h_j|\psi(t)}\neq 0$ only when $j=\{i-1,i,i+1\}$ due to $\ket{\psi(t)}$ not featuring two neighboring sites in the $\ket{-}$ state.    
Working in the thermodynamic limit, we can take each of these cases in turn, finding that 
\begin{equation}
\begin{split}
\braket{\psi(t)|h_i h_i|\psi(t)}=&\frac{1}{81n_{1/2}^4}\left(\gamma(1+e^{i\epsilon t})(1+e^{i\epsilon t})+\gamma(1-e^{i\epsilon t})(1-e^{i\epsilon t})-2\Delta e^{i\epsilon t}  \right)\\
&\times\left(\gamma(1+e^{-i\epsilon t})(1+e^{-i\epsilon t})+\gamma(1-e^{-i\epsilon t})(1-e^{-i\epsilon t})-2\Delta e^{-i\epsilon t} \right)\\
=&\frac{1}{81n_{1/2}^4}\left( 2\gamma(1+e^{2i\epsilon t})-2\Delta e^{i\epsilon t} \right)\left( 2\gamma(1+e^{-2i\epsilon t})-2\Delta e^{-i\epsilon t} \right) \\
=&\frac{16}{81n_{1/2}^4}\left( \gamma \cos(\epsilon t)-\frac{\Delta}{2} \right)^2.
\end{split}
\end{equation}
For the other two cases, we find:
\begin{eqnarray}
\braket{\psi(t)|h_i h_{i+1}|\psi(t)} = \braket{\psi(t)|h_{i+1} h_i|\psi(t)}=\frac{1}{3\sqrt{2}n_{1/2}}\braket{\psi(t)|h_i h_i|\psi(t)} 
=\frac{16}{243\sqrt{2}n_{1/2}^5}\left( \gamma \cos(\epsilon t)-\frac{\Delta}{2} \right)^2.    
\end{eqnarray}
Therefore, the quantum leakage in the thermodynamic limit is
\begin{equation}
\Gamma= \sqrt{N}\sqrt{\frac{16}{81n_{1/2}^4}+2\frac{16}{243\sqrt{2}n_{1/2}^5}}\int^T_0\left|\gamma \cos(\epsilon t)-\frac{\Delta}{2} \right| \; \d t= \frac{4\sqrt{N}}{ 9n_{1/2}^2}\sqrt{1+\frac{\sqrt{2}}{3n_{1/2}}}\int^T_0\left|\gamma \cos(\epsilon t)-\frac{\Delta}{2} \right| \; \d t.
\end{equation}

For finite systems, the same arguments hold. So we get non-zero contributions that are still proportional to $\left|\gamma \cos(\epsilon t)-\frac{\Delta}{2} \right|$. This means that we can get a perfectly periodic orbit with the same driving parameters.  However, the exact expression of $\Gamma$ becomes non-trivial due to the normalization factor.

\subsection{SPT order parameter}

\begin{figure}[htb]
\centering
    \includegraphics[width=0.5\linewidth]{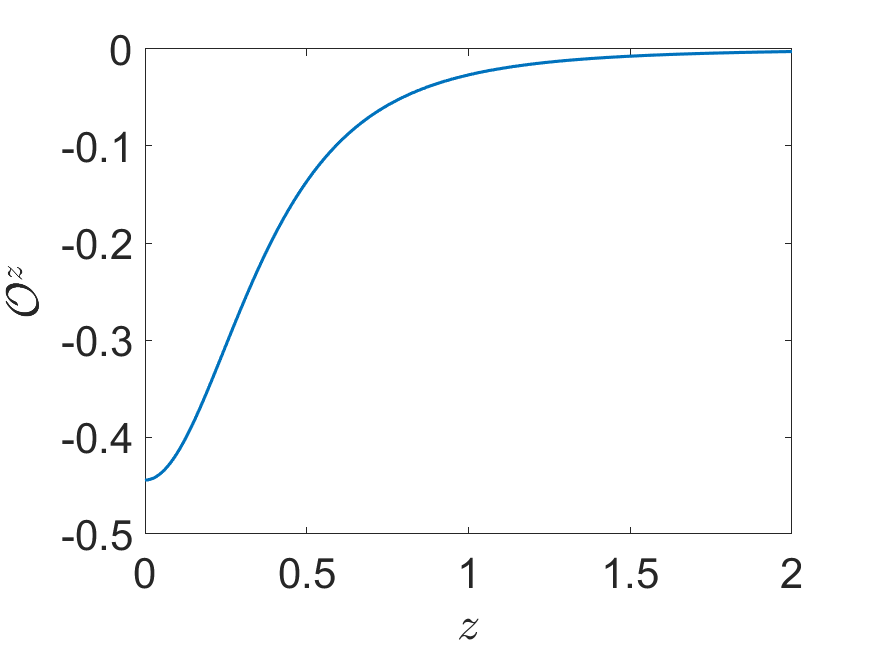}
      \caption{String order of the scarred AKLT initial state in Eq.~(\ref{eq:aklt_init}) as a function of $z$. When $z=0$ the string-order of the AKLT ground-state is recovered.}
         \label{fig:AKLTstring}
\end{figure}

Despite being a finite energy density state, the scarred initial state in Eq.~(\ref{eq:aklt_init}) actually possesses a finite string order for the operator~\cite{Kohmoto1992}
\begin{equation}
\mathcal{O}^z=\lim_{|i-j|\rightarrow \infty}\langle S^z_i \prod^{n=j-1}_{n=i+1}e^{i\pi S^z_n}S^z_j\rangle.
\end{equation}
The behavior of $\mathcal{O}^z$ as a function of $z$ is shown in Fig. \ref{fig:AKLTstring}. When $z=0$ the well-known value of the AKLT string order is recovered, $\mathcal{O}^z=-4/9$. The value of $\mathcal{O}^z$ only depends upon the sign of $z$ and is therefore constant over time during the evolution. It is typically expected that at finite energy density, the string order of an SPT phase is lost. The robustness of the string order in this case is therefore unusual, indicative of the unusual nature of the scarred eigenstates responsible for the periodic trajectory.

\section{Iadecola-Schecter domain-wall preserving model}

Another model with a tower of scarred eigenstates that can be exactly constructed in the Iadecola-Schecter ``domain-wall preserving" spin-1/2 model~\cite{Iadecola2019_3},
\begin{equation}
\hat{\mathcal{H}}_0=\sum_n \lambda(\sigma^x_n-\sigma^z_{n-1}\sigma^x_n\sigma^z_{n+1}) +\Delta\sigma^z_n+J\sigma^z_n \sigma^z_{n+1}.
\end{equation}
The tower of scarred states in this model is generated by the repeated application of the operator
\begin{equation}
Q^+=\sum_n (-1)^n P^0_{n-1}\sigma^+_nP^0_{n+1}
\end{equation}
onto the state $\bigotimes_n \ket{0}$, where $P^0_n=(1-\sigma^z_n)/2$ is the projector onto spin down. These states have energies $E_n=(2\Delta-4J)n+J(N-1)-\Delta N$. While the model overall is unconstrained, the scarred eigenstates obey an emergent kinematic constraint in which no two neighboring sites can be occupied. 

A simple state which exhibits periodic oscillations due to these scarred eigenstates is
\begin{equation}
\ket{\eta}\propto \mathcal{P}_{c} \prod_n[\mathbb{1}+(-1)^n\eta\sigma^+_n]\ket{0}
\end{equation}
where $ \mathcal{P}_{c}=\prod_{n}\left(1-\ket{11}_{n,n+1}\bra{11}_{n,n+1}\right)$ imposes the emergent kinematic constraint. This state can be written simply as a $\chi=2$, $d=2$ MPS where,
w\begin{equation}
A^1=(-1)^n \frac{\eta}{2}\begin{pmatrix}
    -1 & -1 \\
    1 & 1
\end{pmatrix},    \quad
A^0=\begin{pmatrix}
    0 & 0 \\
    -1 & 1
\end{pmatrix}.    
\end{equation}
Choosing $\chi=2$, $d=2$ for our variational manifold, it is straightforward to construct a valid $\hat{\mathcal{H}}_1$. If $\hat{\mathcal{H}}_1\ket{\psi(t)}$ violates the emergent kinematic constraint in at least two places, the state will be annihilated by the tangent space projector. Therefore, it suffices to choose  $\hat{\mathcal{H}}_1$ as
\begin{equation}
\hat{\mathcal{H}}_1=\gamma \sum_n \ket{\Phi}\bra{1,1,1,1} + \mathrm{h.c.}
\end{equation}
The construction of Floquet scars for this model is slightly more complicated than the previous examples. However, we can proceed if we notice that $A^0 A^0 = A^0$ and $A^0 A^1 A^0 = \eta (-1)^n A^0$. Using these relations and fixing $\eta=1$, we introduce the following perturbation
\begin{equation}
\hat{\mathcal{H}}_1=\sum_n  \sigma^+_{n-3}\sigma^+_{n-2}\left[\gamma\sigma^+_{n-1} P^1_n \sigma^+_{n+1}+\Delta(-1)^n\left(\sigma^+_{n-1}\sigma^+_{n}\sigma^+_{n+1}+P^1_{n-1}\sigma^+_{n}P^1_{n+1}\right)\right]\sigma^+_{n+2} \sigma^+_{n+2}+h.c.
\end{equation}
For this perturbation, we find that the instantaneous quantum leakage is proportional to
\begin{equation}
\Gamma \propto \int^T_0  |\gamma+\frac{\Delta}{2}\cos(\epsilon t)| \; \d t
\end{equation}
and thus we obtain perfect Floquet scars when $\gamma=-2\Delta \cos(\epsilon t)$.

\section{The cluster model}

The ZXZ or the cluster model~\cite{ClusterModel} is a spin-$\frac{1}{2}$ chain defined by the Hamiltonian
\begin{equation}
\mathcal{H}_0=-\sum_n\sigma^z_{n-1}\sigma^x_{n}\sigma^z_{n+1}=-\sum_n K^z_{n}. \end{equation}
This model realizes an SPT phase, preserved by a dihedral $\mathbb{Z}_2 \times \mathbb{Z}_2$ symmetry. The dihedral symmetry is generated by spin flips on the odd and even sites of the chain, $U_e=\prod_{n \in even}\sigma^x_n$ and $U_o=\prod_{n \in odd}\sigma^x_n$. Its ground state is the well-known cluster state, widely studied as a resource for quantum computation. The cluster state can be represented as a $\chi=2$ matrix-product state
\begin{equation}
A^\downarrow=\frac{1}{\sqrt{2}}\begin{pmatrix}
0 & 0\\
1 & 1
\end{pmatrix}, \quad
A^\uparrow=\frac{1}{\sqrt{2}}\begin{pmatrix}
-1 & 1\\
0 & 0
\end{pmatrix}.
\end{equation}
The cluster state is an eigenstate of each individual cluster operator $K^z_{n}\ket{\psi}=\ket{\psi}$ with eigenvalue $+1$. The cluster raising/lowering operators can flip the sign of any particular cluster operator, $K^{\pm}_n=\frac{1}{2}(\sigma^z_n+i \sigma^z_{n-1}\sigma^y_{n}\sigma^z_{n+1}$). Using these operators, a complete basis for the spin chain can be constructed, with $K^z_n$ being equivalent to the $\sigma^z_n$ operator in the computational basis. Similarly, $K^x_n=K^{+}_n+K^{-}_n=\sigma^z_n$ and $K^y_n=i(K^{+}_n-K^{-}_n)=\sigma^z_{n-1}\sigma^y_{n}\sigma^z_{n+1}$, are the cluster basis equivalent of $\sigma^x_n$ and $\sigma^y_n$. 

Using these operators it is straightforward to construct a state that oscillates periodically under dynamics generated by the cluster Hamiltonian.  If the state is prepared in the $+1$ eigenstate of each $K^x_n$ operator, it will locally be a superposition of $-1$ and $+1$ eigenstates of $K^z_n$, therefore it will oscillate periodically in the $K^x-K^y$ plane. Since $K^x_n$ is just $\sigma^z_{n}$, initially it is simply $\ket{\psi(t=0)}=\ket{\downarrow \downarrow \downarrow...}$.  The full quantum dynamics can be captured using a $\chi=2$ MPS ansatz by noting that one cluster operator can be flipped from $-1$ to $+1$ by applying applying a $\sigma^z$ to the MPS on the virtual level $A^\sigma_n \rightarrow A^\sigma_n \sigma^z$. Therefore, in terms of MPS tensors the state evolves as:
\begin{figure*}
\centering
    \includegraphics[width=0.32\linewidth]{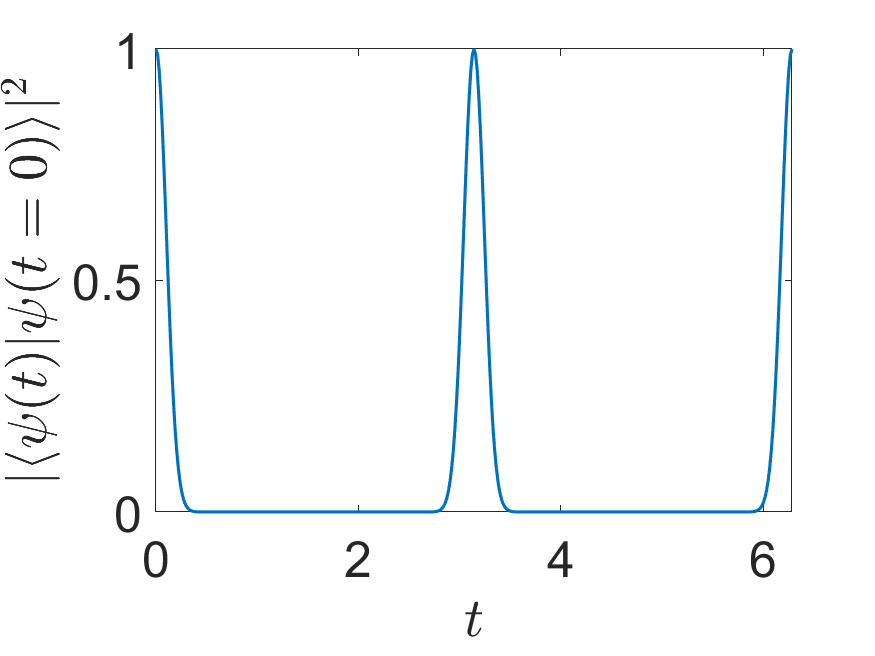}
    \includegraphics[width=0.32\linewidth]{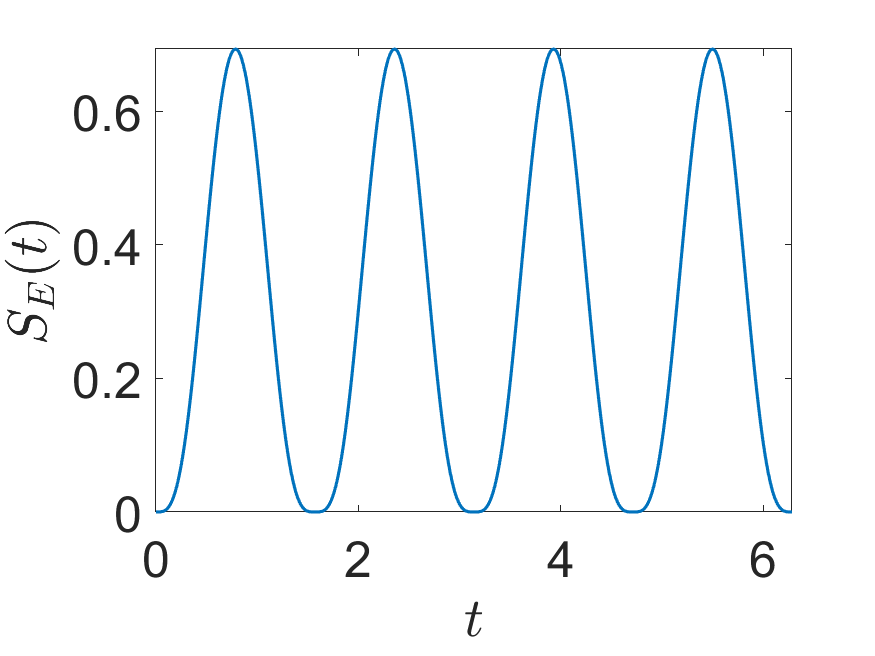}
    \includegraphics[width=0.32\linewidth]
    {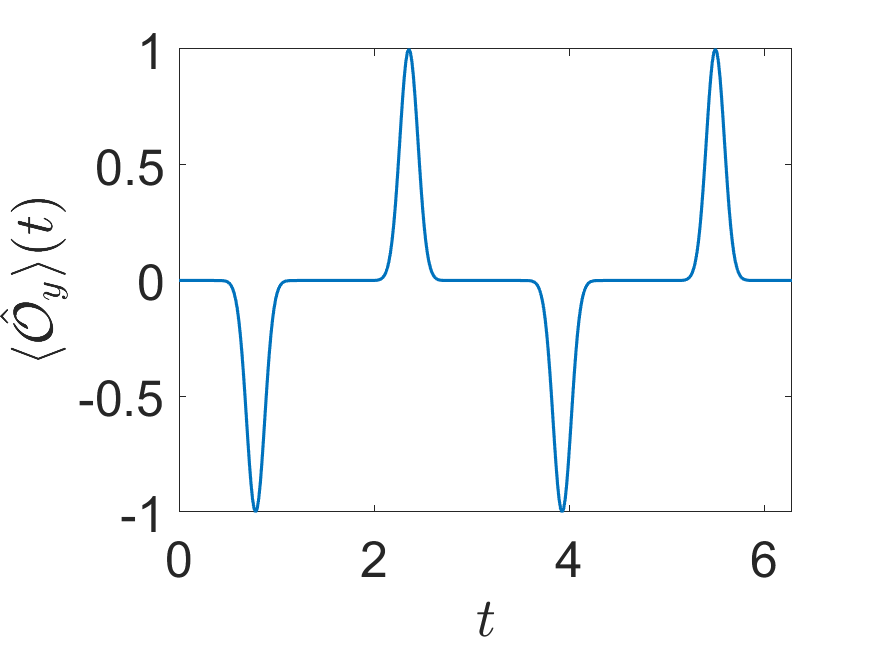}
      \caption{Behavior of the cluster model starting from the initial state $\ket{\psi(t=0)}=\ket{\downarrow \downarrow \downarrow...}$ at $N=50$. Left: Fidelity showing revivals at $t=\pi,2\pi$. Middle: Entanglement entropy over time. Right: String order parameter $\langle \mathcal{\hat{O}}_y\rangle$ with $n_1=10$. $n_2=40$, showing the persistence of SPT order parameter at finite energy density.}
         \label{fig:cluster}
\end{figure*}
\begin{equation}
A^\downarrow(t)=\frac{1}{\sqrt{2}}\begin{pmatrix}
0 & 0\\
1 & 1
\end{pmatrix}\begin{pmatrix}
1+e^{2it} & 0\\
0 & 1-e^{2it}
\end{pmatrix}, \quad
A^\uparrow(t)=\frac{1}{\sqrt{2}}\begin{pmatrix}
-1 & 1\\
0 & 0
\end{pmatrix}\begin{pmatrix}
1+e^{2it} & 0\\
0 & 1-e^{2it}
\end{pmatrix}.
\end{equation}
This state oscillates periodically with period $T=\pi$. When $t=\pi/4$ and $t=3\pi/4$ the state is oriented in the $\pm K^y$ direction. At this point $S_E(t)=\log(2)$ and the state possesses long-range string order such that the operator
\begin{equation}
\mathcal{\hat{O}}_y=\prod^{m=n_2}_{m=n_1}K^y_{m}=(-1)^{(n_1-n_2)}\sigma^z_{n_1-1}\sigma^x_{n_1}\prod^{m^\prime=n_2-1}_{m^\prime=n_1+1}\sigma^y_{m^\prime}\sigma^x_{n_2}\sigma^z_{n_2+1},
\end{equation}
takes the value $\langle \mathcal{\hat{O}}_y\rangle=\pm 1$. In this sense, the state possesses long-range SPT order despite being at finite energy density. The behavior of the fidelity, entanglement entropy and string order parameter can be seen in Fig. \ref{fig:cluster}.

Utilizing the fact that the cluster operators are unitarily equivalent to Pauli operators, we can construct perturbations to $\mathcal{H}_0$ that do not affect the periodic trajectory. It is convenient to work in the cluster basis of eigenstates of $K^z_n$.  Since each site in the cluster basis is oscillating the in the $K_x-K_y$ plane, the state is orthogonal to a singlet constructed in the cluster basis $\ket{ \uparrow\downarrow}_{cluster}-\ket{ \downarrow\uparrow}_{cluster}$. For this reason, the periodic trajectory is unaffected by the addition of terms equivalent to the Heisenberg model in the cluster basis:
\begin{equation}
\hat{\mathcal{H}}_{heis}=J\sum_n K^x_n K^x_{n+1}+K^y_n K^y_{n+1}+K^z_n K^z_{n+1}=J\sum_n \sigma^z_{n-1}(\sigma^x_{n}\sigma^x_{n+1}+\sigma^y_{n}\sigma^y_{n+1})\sigma^z_{n+2}+\sigma^z_{n}\sigma^z_{n+1}.
\end{equation}
We can also introduce driven perturbations to the model that preserve the periodic trajectory. One such perturbation is 
\begin{equation}
\hat{\mathcal{H}}_{1}=\sum_n K^z_n \left(\alpha\left(K^z_{n+2}-K^z_{n+3}\right)+\beta\left(K^z_{n+2}K^x_{n+3}-K^x_{n+2}K^z_{n+3}\right)\right).
\end{equation}
$K^z_n$ flips the cluster on site $n$, and the operators $K^z_{n+2}-K^z_{n+3}$ and $K^z_{n+2}K^x_{n+3}-K^x_{n+2}K^z_{n+3}$ maps the clusters centered on sites $n+2$ and $n+3$ to the cluster singlet, therefore $\hat{\mathcal{H}}_1\ket{\psi(t)}$ is annihilated by the tangent space projector. For this perturbation, we find that the instantaneous quantum leakage is proportional to
\begin{equation}
\Gamma \propto \int^T_0  |\alpha+\beta\cos(2t)| \; \d t,
\end{equation}
and thus we obtain perfect Floquet scars when $\alpha=-\beta \cos(2 t)$. The second term in this perturbation does not commute with $U_e$ or $U_o$ and therefore cannot preserve the SPT order of the cluster model.

\section{Level statistics}

In order to verify that the Floquet models resulting from our procedure are chaotic, we numerically construct the time evolution operator over a full period 
\begin{equation}
\hat{U}_T=\hat{\mathcal{T}}\exp\left[\int_0^T \hat{H}(t)dt \right],
\end{equation}
where $\hat{\mathcal{T}}$ denotes time ordering. The right eigenvectors $\ket{\phi_n}$ of $\hat{U}_T$ are then the Floquet modes with eigenvalues on  the unit circle $e^{i\phi_n}$. If the Floquet operator is chaotic, we expect the spacings between the consecutive $\phi_n$ to obey the prediction of the random matrix theory for the circular orthogonal ensemble (COE). This means that the spacings $s_n=\phi_{n+1}-\phi_{n}$ should obey the Wigner-Dyson distribution~\cite{Mehta2004} and the average spacing ratio $r_n=\min\{s_n/s_{n+1},s_{n+1}/s_n \}$ be close to 0.53 \cite{OganesyanHuse}. Nonetheless, this prediction is applicable only after all the symmetries of $\hat{U}_T$ have been resolved. In this section, we discuss the different models studied in this work and show that they yield spectral statistics in good agreement with COE in all cases.

For the SSH chain, as our initial state of interest is at zero magnetization, $\sum_j \hat{\sigma}^z_j=0$, we restrict to this sector. The only symmetry of the driven model is then spatial inversion for $N$ multiples of 4. However, resolving only this symmetry does not yield the level statistics of a chaotic model. It turns out we also need to consider the action of the spin-flip operator $\hat{X}=\prod_{j=1}^{N} \hat{\sigma}^x_j$. Indeed, let us first split our Hamiltonian between the static and driven parts, which are respectively given by 
\begin{align}
\hat{H}_\mathrm{s}&=\sum_n J_o \sigma^+_{2n+1}\sigma^-_{2n+2}+J_e \sigma^+_{2n+2}\sigma^-_{2n+3}{+} \Delta \sigma^+_{2n+1}\sigma^-_{2n+4}+\mathrm{h.c.}, \\
\hat{H}_\mathrm{d}(t)&=-\frac{i}{2}(J_e+\Delta)\sin(2J_ot) \sum_n(-1)^{n}\left(\sigma^+_{2n+1}\sigma^-_{2n+3} -\sigma^+_{2n+2}\sigma^-_{2n+4}\right)+ \mathrm{h.c.}.
\end{align}
It is then straightforward to check that $\hat{X}$ commutes with $\hat{H}_\mathrm{s}$ but \emph{anticommutes} with $\hat{H}_\mathrm{d}(t)$. The consequences of these relations become clearer when we also notice that $\hat{H}_\mathrm{d}(t+T/2)=-\hat{H}_\mathrm{d}(t)$ due to the driving function being sinusoidal. Let us now introduce the operators $\hat{U}_a$ and $\hat{U}_b$ that describe the evolution for the first and second half of the driving period such that $\hat{U}_T=\hat{U}_b\hat{U}_a$. It is then straightforward to relate them according to
\begin{equation}
\begin{aligned}
\hat{U}_b&=\hat{\mathcal{T}}\exp\left[\int_{T/2}^T \hat{H}(t)dt \right]=\hat{\mathcal{T}}\exp\left[\int_{0}^{T/2} \hat{H}(t+T/2)dt \right]=\hat{\mathcal{T}}\exp\left[\int_{0}^{T/2} \hat{H}_\mathrm{d}-\hat{H}_\mathrm{s}(t)dt \right]\\
&=\hat{X}\hat{X}\hat{\mathcal{T}}\exp\left[\int_{0}^{T/2} \hat{H}_\mathrm{d}-\hat{H}_\mathrm{s}(t)dt \right]=\hat{X}\hat{\mathcal{T}}\exp\left[\int_{0}^{T/2} \hat{H}_\mathrm{d}+\hat{H}_\mathrm{s}(t)dt \right]\hat{X}=\hat{X}\hat{U}_a\hat{X},
\end{aligned}
\end{equation}
where we have used the fact that $\hat{X}^2=\mathbf{1}$. This also means that we can write $\hat{U}_b\hat{X}=\hat{X}\hat{U}_a$, which finally leads to 
\begin{equation}
\hat{U}_T=\hat{U}_b\hat{U}_a=\hat{U}_b\hat{X}\hat{X}\hat{U}_a=(\hat{U}_b\hat{X})(\hat{X}\hat{U}_a)=(\hat{X}\hat{U}_a)^2.
\end{equation}
As a consequence, $\hat{X}\hat{U}_a$ and $\hat{U}_T$ have the same Floquet modes (including the one corresponding to our initial state), but the phase of each mode of $\hat{U}_T$ is twice that of the corresponding $\hat{X}\hat{U}_a$ mode. The computed statistics of the spacings shows a good agreement with COE for $\hat{X}\hat{U}_a$ but not for $\hat{U}_T$. This is a consequence of the mentioned phase doubling. Indeed, as the phases are folded back into the range $[-\pi,\pi]$ this can lead to quasi-degeneracies into the spectrum of $\hat{U}_T$ and deviation from the COE predictions. Nonetheless, $\hat{U}_T$ is still ``chaotic" because it is equivalent to applying the chaotic time-evolution $\hat{X}\hat{U}_a$ operator twice.
\begin{figure*}
\centering
    \includegraphics[width=0.8\linewidth]{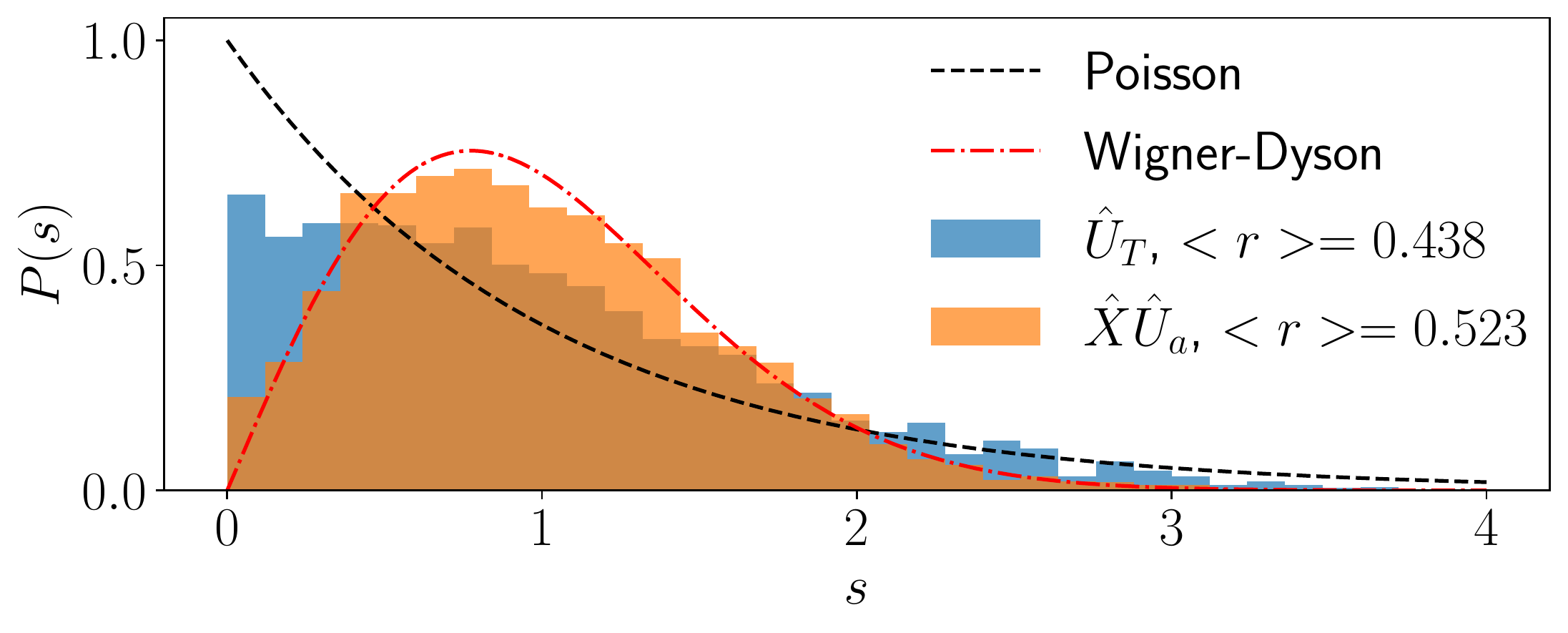}
      \caption{Floquet level spacing statistics for the evolution operators $\hat{U}_T$ and $\hat{X}\hat{U}_a$ for the SSH model at $N=16$. The latter displays good agreement with COE, while the former shows the presence of quasi-degeneracies due to the folding of the spectrum. }
         \label{fig:SSH_stats}
\end{figure*}

Other models studied in this work display a similar relation where a symmetry of the static part of the Hamiltonian anticommutes with the driving term. For the spin-1 XY model with an even number of sites, the only symmetry of the static Hamiltonian is the ``parity'' operator $\hat{Z}_2=\exp \left[ -i\pi \sum_{j=1}^N \hat{S}^z_j \right]$. Meanwhile, the operator $\hat{R}$ that generates spatial inversion commutes with the static terms but anticommutes with the driving term. Using the analogous construction as for the SSH model, we readily obtain the decomposition $\hat{U}_T=(\hat{R}\hat{U}_a)^2$. We checked the level statistics of $\hat{R}\hat{U}_a$ and confirmed it agrees well with the Wigner-Dyson distribution.

The case of the AKLT model is only slightly more complicated. Here, the only symmetry of the full Hamiltonian is also the operator $\hat{Z}_2=\exp \left[ -i\pi \sum_{j=1}^N \hat{S}^z_j \right]$. However the static terms also commute with $\hat{Z}_4=\exp \left[ -i(\pi/2) \sum_{j=1}^N \hat{S}^z_j \right]$, while the driving term anticommutes with it. Let us restrict to a symmetry sector of $\hat{Z}_2$ with eigenvalue $z=\pm 1$. In that sector, we then have $(\hat{Z}_4)^2=\hat{Z}_2=z\mathbf{1}$ and so $\hat{Z}_4=z \hat{Z}_4^\dagger$ as $\hat{Z}_4\hat{Z}_4^\dagger=\mathbf{1}$. We can once again rewrite the full evolution operator $\hat{U}_T=\hat{U}_b\hat{U}_a$ with $\hat{U}_b\hat{Z}_4=\hat{Z}_4\hat{U}_a=z\hat{Z}_4^\dagger\hat{U}_a$ which leads to
\begin{equation}
\hat{U}_T=\hat{U}_b\hat{U}_a=\hat{U}_b\hat{Z}_4\hat{Z}_4^\dagger\hat{U}_a=z\hat{U}_b\hat{Z}_4\hat{Z}_4\hat{U}_a=z(\hat{Z}_4\hat{U}_a)^2.
\end{equation}
As $z=\pm 1$, it does not stretch the spectrum and lead to folding. So, in the end, we recover the same construction as for the SSH, and we find good agreement between the spectral statistics of ${Z}_4\hat{U}_a$ and the COE prediction, showing that our Floquet operator $\hat{U}_T$ is indeed chaotic. 

Despite the fact that in all three models considered the Floquet evolution operator can be written as a square of some operator, this is not necessary for our construction. As an example, we note that we can change the driving term in the AKLT case to 
\begin{equation}
\hat{H}_\mathrm{d}(t)=2\gamma\cos{(\epsilon t)}\left[\kappa\sum_n \hat{P}^-_n \hat{P}^-_{n+1} +\sum_n (-1)^n\Delta\ket{0,+,0,+}\bra{\chi_{-}}+\mathrm{h.c.}\right],
\end{equation}
where $\hat{P}^-=\ket{-}\bra{-}$ is the projector on the $-1$ spin state and $\kappa$ is a free parameter. The $\hat{P}^-_n \hat{P}^-_{n+1}$ perturbation does not affect the periodic trajectory as the latter has no overlap on configurations with neighbouring $-1$. However, as $\hat{P}^-_n \hat{P}^-_{n+1}$ commutes with $\hat{Z}_4$, it means that $\hat{H}_\mathrm{d}(t)$ no longer anticommutes with this operator. As a consequence, $\hat{U}_T$ does not factorize and we indeed find it has chaotic spectral statistics. More generally, we can notice that the factorization found in these models also relies on the condition  $\hat{H}_\mathrm{d}(t+T/2)=\; -\hat{H}_\mathrm{d}(t)$. So we do not expect it to appear for more complex trajectories, where the driving should not simply be sine or cosine.

\end{document}